\documentclass[12pt]{article}
\pdfoutput=1

\usepackage{jheppub}
\usepackage{lipsum}
\usepackage{braket}
\usepackage{tensor}
\usepackage{dsfont}
\usepackage{tcolorbox}
\usepackage{caption}
\usepackage{subcaption}
\usepackage{tikz}
\usetikzlibrary{decorations.pathreplacing,calligraphy}
\usetikzlibrary{decorations.pathreplacing}
\usetikzlibrary{decorations.pathmorphing, patterns,shapes}
\usepackage{bm}
\usepackage{caption}
\usepackage{multirow}
\usepackage{subcaption}
\usepackage{ragged2e}
\usepackage{mathtools}
\usepackage{cancel}
\usepackage{comment}
\usepackage{xcolor}
\DeclareCaptionJustification{justified}{\justifying}
\usepackage{hyperref}
\usepackage{amssymb}
\usepackage{pgfplots}

\usepackage{tikz}
\usetikzlibrary{arrows,decorations.markings,shapes.geometric,decorations.pathmorphing}
\tikzset{snake it/.style={decorate, decoration=snake}}
\usetikzlibrary{decorations.pathreplacing,calligraphy}
\usetikzlibrary{decorations.pathreplacing}
\usetikzlibrary{decorations.pathmorphing, patterns,shapes}
\usepackage{bm}

\newcommand{\op}[1]{\boldsymbol{#1}}
\newcommand{\de}{\mathrm d}
\newcommand{\sumint}{\ \mathclap{\displaystyle\int}\mathclap{\textstyle\sum}\ \ }

\pgfdeclareverticalshading{rainbow}{100bp}
{color(0bp)=(red); color(25bp)=(red); color(35bp)=(yellow);
color(45bp)=(green); color(55bp)=(cyan); color(65bp)=(blue);
color(75bp)=(violet); color(100bp)=(violet)}

\begin{document}

\vspace*{-.6in} \thispagestyle{empty}
\begin{flushright}
{\tt DESY-24-136}\\
{\tt ZMP-HH/24-20}
\end{flushright}
\vspace{1cm} {\Large
\begin{center}
{\bf Constraints on RG Flows from Protected Operators}\\
\end{center}}
\vspace{1cm}
\begin{center}
{\bf Florent Baume$^{a}\footnote{florent.baume@desy.de}$, Alessio Miscioscia$^b\footnote{alessio.miscioscia@desy.de}$ and Elli Pomoni$^{b}\footnote{elli.pomoni@desy.de}$}\\[2cm] 
{
$^{a}$ II. Institut f\"ur  theoretische Physik, Universit\"at Hamburg, Luruper Chaussee 149, 22607 Hamburg, Germany\\
$^{b}$ Deutsches Elektronen-Synchrotron DESY, Notkestr. 85, 22607 Hamburg, Germany\\
}
\vspace{1cm}
\end{center}

\vspace{4mm}

\begin{abstract}

\ We consider
protected operators with the same conformal dimensions in the ultraviolet and infrared fixed point.
We derive a sum rule for the difference between the two-point function coefficient of these operators in the ultraviolet and infrared fixed point which depends on the two-point function of the scalar operator.
In even dimensional conformal field theories, scalar operators with exactly integer conformal dimensions are associated with Type-B conformal anomalies. The sum rule, in these cases, computes differences between Type-B anomaly coefficients. We argue the positivity of this difference in cases in which the conformal manifold contains weakly coupled theories. 
The results are tested in free theories as well as in $\mathcal N = 2$ superconformal QCD, necklace quivers and holographic RG flows.
We further derive sum rules for currents and stress tensor two-point functions.

 \end{abstract}
\vspace{.2in}
\vspace{.3in}

\newpage

{
\tableofcontents
}
\newpage

\setlength{\parskip}{0.1in}

\section{Introduction}\label{sec:introduction}

Renormalization Group (RG) flows are one of the most important and fascinating
topics in theoretical physics, providing both conceptual and concrete
connections between the long-distance, macroscopic behavior of a physical
system and its short-distance, microscopic description \cite{Wilson:1971bg,
Wilson:1971dh, Wilson:1973jj, Wilson:1974mb, Polchinski:1983gv}. The concept of
\textit{irreversiblity} of RG flows was first shown in two-dimensional Quantum
Field Theory (QFT) by Zamolodchikov \cite{Zamolodchikov:1986gt}, where he
demonstrated the existence of a function monotonically decreasing along RG
flows. This so-called $C$\emph{-function} can then be interpreted as a
non-perturbative counting of degrees of freedom. Similar properties have been
established in three \cite{Klebanov:2011gs} and four dimensions
\cite{Komargodski:2011vj} and are referred to as the $F$- and $a$-theorems,
respectively. Cardy further proposed that in even dimensions, the relevant $C$-function
is related to the trace anomaly \cite{Cardy:1988cwa}. Attempts have since been
made to generalise the two- and four-dimensional proofs to six dimensions and
higher using background dilatons \cite{Elvang:2012st, Elvang:2012yc,
Baume:2013ika, Stergiou:2016uqq}, but a general proof remains
elusive.\footnote{It has however been established and checked for large
classes of RG flows between superconformal field theories (SCFTs)
\cite{Heckman:2015axa, Cordova:2015fha, Mekareeya:2016yal,
Cordova:2020tij, Heckman:2021nwg, Baume:2023onr, Fazzi:2023ulb}.} 

Further generalizations to non-unitary two-dimensional QFTs
\cite{Castro-Alvaredo:2017udm} or including defects \cite{Cuomo:2021rkm,
Casini:2022bsu, Casini:2023kyj}, as well as alternative proofs highlighting
connections with fascinating non-perturbative results \cite{Cappelli:1990yc,
Hartman:2023ccw, Hartman:2023qdn} have also been discussed in the literature.
These $C$-functions being deeply related to the stress tensor, it is
natural to ask whether similar quantities can be associated with other
particular operators that can be tracked along an RG flow, and the constraints
they impose. Such questions have been explored for instance in two dimensions
for flavor currents \cite{Vilasis-Cardona:1994oke}, and to study critical
exponents \cite{Delfino:1996nf}.

In this work, we explore the evolution of certain scalar operators
along RG flows, whose two-point function in the fixed points is constrained to
take the form
\begin{equation}
    \langle \mathcal O(x) \overline{\mathcal O}(0) \rangle_\text{UV} = \frac{C^\text{UV}_\Delta}{x^{2\Delta}} \ ,\qquad\qquad
    \langle \mathcal O(x) \overline{\mathcal O}(0) \rangle_\text{IR} = \frac{C^\text{IR}_\Delta}{x^{2\Delta}} \ ,
\end{equation}
where $\Delta$ is the conformal dimension of the operator $\mathcal O$, and the
numerators $C_\text{UV}$, $C_\text{IR}$ are positive numbers.\footnote{The
positivity of the two-point functions at the fixed point is a
consequence of unitarity. Furthermore, note that we are not excluding
the case for which $C_\Delta^\text{IR} = 0$, occurring when the operator is completely
integrated out in the deep IR.} We stress that the fact that the conformal
dimensions in the UV and the IR are the same is a consequence of the assumption
that the operator is protected, as discussed in more detail in Section
\ref{sec:sumrule}. The presence of such operators in a generic
QFT is of course no guaranteed, and it is often the consequence of an unbroken
symmetry. They are for instance quite common supersymmetric theories, where
certain BPS conditions forbid mixing along the RG flow, the most obvious example being operators that are part of the chiral ring in $4d$ $\mathcal{N}=1$ theories.
In the appropriate choice of normalisation the coefficients $C_\Delta^\text{UV}$
and $C_\Delta^\text{IR}$ then encode information about the theory and capture part of the data of their chiral ring.

We will however assume the existence of these operators without making any
particular requirement about the origin of the underlying protection mechanism.
In particular, we will not assume any supersymmetry, our goal rather being to
derive general properties of RG flows involving this type of operators.  We
will then be able to derive a sum rule for the difference $\delta C_\Delta$ of
the numerators for any protected operator $\mathcal{O}$ whose scaling dimension
$\Delta$ remains constant along the RG flow:
\begin{equation}\label{eq:sr}
    \delta C_{\Delta} = C_\Delta^\text{UV} - C_\Delta^\text{IR} = 
	\int \de^d x\,  \mathcal{D}_\Delta \langle \mathcal O(x) \overline{\mathcal O}(0)\rangle\,,
\end{equation}
where the differential operator $\mathcal{D}_\Delta$ depends on the scaling
dimension, and whose exact form is derived in Section
\ref{sec:derivationsumrule}. Similar sum rules have been crucial in deriving
the $c$-theorem and its generalizations \cite{Cardy:1988tj, Schwimmer:2010za,
Hartman:2023ccw, Hartman:2023qdn, Hartman:2024xkw, Karateev:2023mrb}. 
As we will see in our case the differential operators $\mathcal{D}_\Delta$ is not
manifestly positive. Furthermore, the two-point function is in general very
difficult to compute non-perturbatively off criticality. 

In order to circumvent this issue, we will expand this correlator in terms of
its \textit{spectral decomposition} as a sum of form factors $\langle 0|
\mathcal O |\alpha\rangle$, where $\ket{\alpha}$ is a state in the Fock space
defined at large distances. In particular, we show that the sum rule given in
equation \eqref{eq:sr} is not sensitive to single-particle states. As a
consequence, it receives contributions only from multi-particle states. We then
explore cases in which in the IR theory the operator $\mathcal{O}$ vanishes,
and comment on the general case by using perturbative expansions for the
spectral decomposition around the high-energy limit. 

Our results are particularly interesting for theories in even spacetime
dimensions where the protected operators have integer conformal dimensions
$\Delta\in \mathbb{N}$. 
There, the relevant two-point function coefficients
are associated with conformal anomalies \cite{Bzowski:2013sza,
Bzowski:2015pba}, and have in particular been discussed in the context of the
supersymmetric theories \cite{Niarchos:2019onf, Niarchos:2020nxk,
		Andriolo:2022lcb, Gomis:2015yaa, Schwimmer:2018hdl, Kuzenko:2019vvi,
Schwimmer:2023nzk}. By combining our results with those of references
\cite{Niarchos:2019onf, Niarchos:2020nxk} we will argue that $\delta C_\Delta
\ge 0$ in the case of Type-B anomalies associated with theories whose conformal
manifold contains a free point. Our strategy will be to use the fact that
$C_\Delta$ is covariantly constant over along conformal manifolds
\cite{Andriolo:2022lcb} to tackle the difference $\delta C_\Delta$ at weak
coupling, where non-perturbative effects are suppressed.

The rest of this work is organized as follows:
\begin{itemize}
	\item[$\star$] In Section \ref{sec:sumrule} we present a derivation of the
			sum rule resembling the case of the $c$-function
			\cite{Zamolodchikov:1986gt,Cardy:1988tj}, with the crucial difference that a
			non-trivial differential operator is applied to the two-point
			function of the protected operators. We further adapt the derivation to study two-point functions of currents and the stress tensor.
    
	\item[$\star$] In Section \ref{sec:sumruleandspectral}, after briefly
			reviewing the spectral decomposition of CFTs, we combine the sum
			rule with the spectral decomposition to conclude that
			single-particle states do not contribute. Restricting ourselves to
			even spacetime dimensions and integer conformal dimensions of the
			protected operators, we then discuss our result in relation to
			Type-B anomalies. In cases where those anomalies are associated to
			theories whose UV CFT admits a free-field limit, we argue that $\delta
			C_\Delta \ge 0$. We also comment on the connection between our
			results and the average null energy operator.
    
	\item[$\star$] In Section \ref{sec:examples} we apply the sum rule to
			explicit examples, namely free theories, supersymmetric flows from
			$\mathcal N = 2$ SQCD and quiver theories, and holographic RG
			flows.
\end{itemize}
We discuss open questions and give our conclusions in Section
\ref{sec:conclusions}. In addition, Appendix \ref{sec:CPToff-critical} gives
further details on the conformal perturbation theory expansion necessary to
complete the proof and probe the convergence of the sum rule in Section \ref{sec:sumrule}, and Appendix
\ref{sec:spectral-decomposition} reviews some aspects of the spectral
decomposition used in the work.

\section{The Sum Rule}\label{sec:sumrule}

Let us derive a sum rule for the evolution of the coefficients of the two-point
function of protected operators. Our setup is the following: we consider an
ultraviolet (UV) conformal field theory (CFT) which is deformed by a relevant
operator, triggering an RG flow.  Even though we will not assume the existence
of a Lagrangian description, it is useful to consider a formal action for the
UV CFT, $\mathcal A_{UV}$, deformed by a set of relevant operators $\Phi^I$:
\begin{equation}\label{eq:formalAction}
		\mathcal A = \mathcal A_{UV} + \int \de^d x \, g_I\, \Phi^I(x) \ .
\end{equation}

In order to avoid an old---but still active---discussion on the nature of the
endpoints of RG flows \cite{Polchinski:1987dy, Riva:2005gd, Luty:2012ww,
Dymarsky:2013pqa, Baume:2014rla, Dorigoni:2009ra, Gimenez-Grau:2023lpz}, we
assume that the deep infrared (IR) is described by either a trivial theory with
no local degrees of freedom or another CFT. We further assume that the spectrum of
the UV theory contains an operator, $\mathcal O$, that is protected along the
flow and does not mix with other operators, meaning that it has the same conformal dimension in the UV and IR fixed point. In practice, this means that along
the RG flow parameterized by a scale $\Lambda$, the connected component of the two-point function
of $\mathcal{O}$ at separated points is given by
\begin{equation}\label{eq:offctp}
    \langle \mathcal O(x) \overline{\mathcal O}(0) \rangle = \frac{C_\Delta(\Lambda |x|)}{|x|^{2\Delta}} \,,
\end{equation}
where $\Delta $ is the conformal dimension of the operator $\mathcal O$ at both
the UV and IR fixed point. The fact that the operator is protected can be
understood from the fact that the function $C_\Delta$ asymptotes to a constant
in both the UV the IR:
\begin{equation}\label{CUVIR-def}
		C_{\Delta}^\text{UV}=\lim_{|x|\to0}C_{\Delta}(\Lambda |x|)\,,\qquad
		C_{\Delta}^\text{IR}=\lim_{|x|\to\infty}C_{\Delta}(\Lambda |x|)\,.
\end{equation}
We stress again that this is not a generic situation: in the most general case
the conformal dimension in the UV is different from the conformal dimension in
the IR. This feature is common in supersymmetric theories, where these types of
protected operator correspond to for instance chiral-ring operators.
There, they are protected by BPS condition interpreted as null vectors in the
Hilbert space, ensuring that their two-point function is of the form given in
equation \eqref{eq:offctp}. In the sequel, we will however not restrict
ourselves to supersymmetric theories, and only assume the existence of an
operator with the properties described above.

Furthermore, note that the fact that the operator is encoded in the behavior of
the numerator of its two-point function is also the case for conserved currents
such as the stress tensor or flavor currents. In those cases, this is ensured
by the conservation equation. The evolution of the equivalent of $C_\Delta$ has
been extensively studied \cite{Zamolodchikov:1986gt, Castro-Alvaredo:2017udm,
Hartman:2023ccw, Cappelli:1990yc, Vilasis-Cardona:1994oke, Karateev:2019ymz},
and we will derive similar sum rules for those cases as well.

\subsection{Derivation of the Sum Rule}\label{sec:derivationsumrule}

Our strategy to derive the sum rule is similar to what was done in the most common
derivation of the $c$-theorem in two dimensions \cite{Zamolodchikov:1986gt,
Cardy:1988tj}. There, the derivative of the $c$-function, interpolating the
central charges of UV and IR theories, was written in terms of the two-point
function of the trace of the stress tensor. Integrating the resulting
expression, one obtains a sum rule for the difference $\delta c =
c^\text{UV}-c^\text{IR}$. We will follow similar procedure for protected
operators as we are interested in the difference of the quantities defined in
equation \eqref{CUVIR-def}:
\begin{equation}
		\delta C_{\Delta} = C_\Delta^\text{UV}-C_{\Delta}^\text{IR}\,.
\end{equation}
We therefore start with the two-point function of a protected operator
$\mathcal{O}$ defined at any point of the RG flow, as shown in equation
\eqref{eq:offctp}. Applying the d'Alembertian to the correlator, for
any spacetime dimension $d$ we have the relation:
\begin{equation}\label{box-2-pt}
		\left(x^2\Box-  4\Delta\left(\Delta - \frac{d-2}{2}\right) \right)\langle \mathcal O(x)\overline{\mathcal O}(0)\rangle = 2(d-4 \Delta) \frac{C_{\Delta}'}{|x|^{2(\Delta-1)}}+ 4 \frac{C_{\Delta}''}{|x|^{2(\Delta-2)}} \ ,
\end{equation}
where primed quantities indicate partial derivation with respect to $|x|^2$:
$C^\prime_\Delta = \frac{\partial C_\Delta}{\partial |x|^2}$. This relation will enable
us to extract a sum rule for $\delta C_{\Delta}$. Indeed, focusing first on the
right-hand side of equation \eqref{box-2-pt}, by integrating over $|x|^2$ we find that 
\begin{equation}\label{sum-rule-with-boundary}
	\int_0^\infty d |x|^2 \left(2(d-4 \Delta) C_{\Delta}' + 4 |x|^2 C_{\Delta}''\right) = 
    8\left(\Delta-\frac{d-2}{4}\right)\,\delta C_\Delta + 4 \left. \left (|x|^2 \frac{\partial}{\partial |x|^2}C_{\Delta}\right) \right |_{|x|^2 = 0}^{|x|^2 = \infty}\,,
\end{equation}
where we have used integration by parts and equation \eqref{CUVIR-def}. The
boundary term on the right-hand side of equation \eqref{sum-rule-with-boundary}
can be shown to always vanish by using perturbation theory around both ends of
the RG flow. Using the formal action given in equation \eqref{eq:formalAction},
we have the following expansion around the UV fixed point 
\begin{equation}\label{eq:CPTUV}
		C_\Delta = C^\text{UV}_\Delta +\sum_{I} g_I c_1^I\, |x|^{d-\Delta_{\Phi^I}}+\ldots\ .
\end{equation}
A derivation of this expansion, including higher-order terms, is given in
Appendix \ref{sec:CPToff-critical}. The coefficients $c_{1}^I$ can be computed
in terms of the UV conformal data, but their precise value will not be relevant
for our purpose. As the deformation operators $\Phi^I$ are by assumption
relevant in order to trigger a non-trivial RG flow, we have
$d-\Delta_{\Phi^I}>0$, from which we infer that the boundary contribution in
the UV is trivial:
\begin{equation}
    \lim_{|x|^2 \to 0} |x|^2 \frac{\partial C_\Delta}{\partial |x|^2}  =\lim_{|x|^2 \to 0} \sum_{I}\left[\frac{d-\Delta_{\Phi^I}}{2} c_1^I |x|^{d-\Delta_{\Phi^I}} +\ldots \right] = 0 \ ,
\end{equation}
A similar analysis can be performed around the deep infrared. At the end of the
flow, we perturb the CFT with irrelevant operators $\Psi^I$ with coupling
constants $\lambda_I$. We then obtain a similar (conformal) perturbative
expansion:
\begin{equation}\label{eq:CPTIR}
		C_\Delta = C^\text{IR}_\Delta + \sum_{I}\tilde{c}_1^I \lambda |x|^{d-\Delta_{\Psi^I}} + \ldots \ .
\end{equation}
The operator at the end of the RG flow being irrelevant it has conformal
dimension $d-\Delta_{\Psi^I}<0$, and we therefore have
\begin{equation}
	\lim_{|x|^2 \to \infty}  |x|^2 \frac{\partial}{\partial |x|^2} C_\Delta  =\lim_{|x|^2 \to \infty} \left[\frac{d-\Delta_{\Psi^I}}{2} \tilde c_1^I |x|^{d-\Delta_{\Psi^I}} +\ldots \right] = 0 \,.
\end{equation}
We defer to Appendix \ref{sec:CPToff-critical} for additional details. We 
can therefore conclude that the last term in equation
\eqref{sum-rule-with-boundary} vanishes, so that $\delta C_\Delta$ can be
expressed as the integral of a differential operator acting on the two-point
function, as advertised around equation \eqref{eq:sr}. Restoring the numerical
factors coming from the change of variables to obtain an integral over $|x|^2$,
we find:
\begin{equation}\label{sum-rule-position-space}
 \delta C_\Delta =\frac{\Gamma(\frac{d}{2})}{4\pi^{\frac{d}{2}}(2\Delta-\nu)}
	\int \de^d x \ |x|^{2\Delta-d} \left(|x|^2 \Box - 4\Delta\left (\Delta - \nu\right)\right) \langle \mathcal O(x) \overline{\mathcal O}(0)\rangle\,,
\end{equation}
where we defined $\nu=\frac{1}{2}(d-2)$, the unitarity bound for the conformal
dimension of scalar operators---saturated by free scalar fields. 

The convergence of this sum rule is guaranteed by conformal perturbation
theory. In fact, around the UV and the IR fixed points where singularities are
in general present, the expansions in equations \eqref{eq:CPTUV} and
\eqref{eq:CPTIR} ensure that the integral in the sum rule is well behaved, and
that the sum rule is convergent. 

In the next section, we will show that this sum rule can be used to find bounds
on $\delta C_\Delta$, and we check its validity explicitly in a variety of
examples in Section \ref{sec:examples}. 

\subsection{Conserved Currents and the Stress Tensor}
In the derivation of the sum rule given in equation
\eqref{sum-rule-position-space}, we have only made use of the functional form
of the two-point function of the protected operator, see equation
\eqref{eq:offctp}. While the focus of this work is on scalar operators, this
nonetheless enable us to extend it to the case of other spinning operators that
are under control under the RG flow, namely the stress tensor and conserved
currents associated with unbroken flavor symmetries.

\paragraph{Conserved Currents:} the conservation equation for unbroken symmetries can be
used to track conserved currents and safely define their two-point function
along the RG flow. We will focus here on the parity-even component of the
correlator of currents associated with an Abelian symmetry. The non-Abelian
case can similarly be obtained by choosing an orthogonal Killing basis for the
generators of adjoint representation: $\operatorname{Tr}(T^aT^b)\propto \delta^{ab}$.

A Lorentz-scalar correlator can then be obtained by contracting the indices of the
two currents, in which case we obtain a two-point function similar to that of
protected operators \cite{Karateev:2020axc}:\footnote{The uncontracted
two-point function $\left<J^\mu J^\nu\right>$ depends generically on two
independent functions, $C_J^{(1)}(\Lambda|x|)$ and $C_J^{(2)}(\Lambda|x|)$, which gives the
flavor central charge $C_J$ at the CFT points. We
defer to Appendix A of reference \cite{Karateev:2020axc} for a derivation of
equation \eqref{two-pt-currents}.}
\begin{equation}\label{two-pt-currents}
		\langle J^\mu (x) J_\mu (0) \rangle = (d-2)\frac{ C_J(\Lambda |x|)}{|x|^{2(d-1)}}\,.
\end{equation}
At the two fixed points, the function $C_J(\Lambda|x|)$ reduces to the flavor
central charge of the two CFTs, which is a positive number by unitarity:
\begin{equation}
		C^\text{UV}_J =	\lim_{|x|\to 0}\,C_J(\Lambda |x|) \,,\qquad
		C_J^\text{IR} =	\lim_{|x|\to \infty}C_J(\Lambda |x|)\,.
\end{equation}
We will now assume $d>2$ to find a sum rule for $\delta C_J = C_J^\text{UV}
- C_J^\text{IR}$. As the functional form of equation \eqref{two-pt-currents} is the same as that of protected scalar operators, we can follow the same procedure procedure to find:
\begin{equation}
    \delta C_J =\frac{\Gamma(\frac{d}{2})}{8\pi^{\frac{d}{2}}(3d-2) (d-2)}
	\int \de^d x \ |x|^{d-2} \left(|x|^2 \Box - 2 d (d-1)\right) \langle J^\mu(x) J_\mu (0)\rangle \ ,
\end{equation}
As for that of scalar operators, this sum rule is not manifestly positive
definite. A similar sum rule for the flavor central charge was recently
proposed in reference \cite{Karateev:2020axc}. There, the sum rule is a
consequence of the conservation equation. It would be interesting to understand
the connections between these sum rules, and check if they differ or are
simply equivalent. We leave a detailed study for future work. In the case of $d
= 2$, positivity of $\delta C_J$ was proven in reference
\cite{Vilasis-Cardona:1994oke}.

\paragraph{The stress tensor:} we can repeat the same reasoning to the
parity-even sector of the two-point function of the stress tensor. Contracting
the indices, we obtain \cite{Karateev:2020axc}
\begin{equation}\label{two-pt-stress}
		\langle T^{\mu \nu}(x) T_{\mu \nu}\rangle = \frac{(d-1)(d-2)}{2}\frac{C_T(\Lambda |x|)}{|x|^{2d}}  \,.
\end{equation}
where we find once again a correlator of the form given in equation
\eqref{two-pt-currents} for protected scalar operators. At the two fixed
points we have:
\begin{equation}
		C^\text{UV}_T = \lim_{|x| \to 0} C_T(\Lambda |x|) \,,\qquad
	C^\text{IR}_T = \lim_{|x| \to \infty} C_T(\Lambda |x|)  \ ,
\end{equation}
so that $C^T(\Lambda |x|)$ interpolates between the two central charges of the
UV and IR CFTs. In two dimension $C_T(\Lambda|x|)$ is a $C$-function, but we
cannot find a sum rule in that case due to the factor $d-2$ in equation
\eqref{two-pt-stress}. When $d>2$, we can again use the procedure for scalar
protected operators described above, and we conclude that:
\begin{equation}
    \delta C_T =\frac{\Gamma(\frac{d}{2})}{4\pi^{\frac{d}{2}}(3d+2) (d-2) (d-1)}
	\int \de^d x \ |x|^{d} \left(|x|^2 \Box - 2 d (d+2)\right) \langle T^{\mu \nu}(x) T_{\mu \nu} (0)\rangle\,.
\end{equation}
This equation is very different from the sum rule given in reference
\cite{Karateev:2020axc} for the same quantity (see equations (2.31) and
(2.32) therein). It would be interesting to compare the two sum rules, and check whether
they are equivalent or encode different properties. We leave this analysis for
future work. Furthermore notice that, as for flavor currents, a theorem on the
positivity of $\delta C_T$ was proven only in two dimensions
\cite{Zamolodchikov:1986gt}. The fact that the case $d=2$ is behaves
differently than in higher dimension is consistent with the non-positivity of
the integrand in the sum rule above. The study of the evolution of the the
central charge is nonetheless a current subject of research
\cite{Karateev:2023mrb,Hartman:2024xkw,Karateev:2024skm}. Even if a positivity
theorem cannot be constructed as counterexamples are known
\cite{Nakayama:2012jv,Anselmi:1997am}, it is however interesting to bound the
difference $\delta C_T$ and describe it in terms of physical quantities via sum
rules.

All the sum rules derived above are not manifestly positive due to the presence
of the differential operator applied to the two-point function, which is itself
not manifestly positive definite. As shown in references \cite{Karateev:2023mrb,
Hartman:2024xkw} it is still possible to study the resulting sum rule. In the
following section we will do that for the scalar case by combining the sum rule
with the spectral decomposition of the two-point function.

\section{Spectral Decomposition and Constraints on RG Flows}\label{sec:sumruleandspectral}

The sum rule we have derived in equation \eqref{sum-rule-position-space} give
constraints on the possible RG flows. They can however be difficult to extract,
as the two-point function, in general, involves non-perturbative effects, and
makes its evaluation away from the fixed points difficult, as conformal
perturbation theory cannot be reliably used.

In order to nonetheless find constraints on the RG flows from the protected
operators, we will combine the sum rule with the spectral decomposition of a
scalar two-point function. The latter is an expansion in terms of form factors
$\left |\langle 0  | \mathcal O | \alpha\rangle\right|$, where the set of
states $\ket \alpha$ represent a basis of the asymptotic Fock basis. As briefly
reviewed in Appendix \ref{sec:spectral-decomposition}, applied to the two-point
function of scalar operators, it takes the form:
\begin{equation}\label{spectral-decomposition-Gs}
    \langle \mathcal O(x) \mathcal O(0)\rangle = \int_0^\infty ds \ \rho(s) G_{s}(x) \ , 
\end{equation}
where  $G_s(x)$ is the propagator of a free scalar field of mass  $m^2 = s$ in $d$ dimension
\begin{equation}\label{free-propagator}
		G_s(x) = \int \frac{d^d p}{(2 \pi)^d} \frac{e^{i p x}}{p^2+s} = \frac{1}{(2 \pi)^{\nu+1}} \left(\frac{\sqrt{s}}{|x|}\right)^{\nu} \mathrm K_{\nu}( \sqrt{s}x) \,,\qquad
		\nu = \frac{d-2}{2}\,.
\end{equation}
which is written in terms of the modified Bessel function of the second kind
$K_\alpha(x)$.

Even though the spectral density $\rho(s)$ is arguably the simplest
non-perturbative quantity in Quantum Field Theory, its explicit expression is
known in very few cases. Recent attempts have been made to constrain the
spectral density at the non-perturbative level using modern methods of the conformal bootstrap \cite{Karateev:2019ymz, Karateev:2020axc, Cuomo:2024vfk}.
We will here instead make use of only some of its analytical properties and
asymptotic behavior.

\subsection{The Spectral Decomposition in CFTs}\label{sec:Anomaliesfromsd}

Before combining the spectral decomposition with the sum rule we have found in
the previous section to get new constraints on the RG flows, we will examine
some of its intriguing properties in the case of CFTs, particularly its
connection to conformal anomalies. 

In the context of the CFTs, the spectral decomposition of the stress tensor has
been discussed in reference \cite{Cappelli:1990yc} and more recently in
references \cite{Karateev:2019ymz, Karateev:2020axc}. We provide a brief
overview of their results in Appendix \ref{sec:stressTensorCFT}. By adapting
their methods in the case of protected scalar operators, we will show that
their spectral decomposition is sufficient to reproduce the structure of the
associated conformal anomalies.

Contrary to the case of the stress tensor, the absence of a Lorentz structure
for scalars implies that only spin-zero states can contribute. Furthermore,
imposing scale invariance, we are left with only two possible contributions for
the spectral density $\rho(s)$:\footnote{Observe that $\widetilde C_\Delta
\propto C_\Delta$, and the two constant differs only by factors depending on the spacetime
dimension and conformal dimension of the operator.}
\begin{equation}\label{eq:CFTspectraldensity}
\text{a)} \quad \rho(s) = \widetilde C_\Delta \, s^{\Delta-\frac{d-2}{2}} \delta(s) \,,\qquad\text{or}\qquad
	\text{b)} \quad \rho(s) = \widetilde C_\Delta \, s^{\Delta-\frac{d}{2}} \ .
\end{equation}
The first possibility leads to a vanishing correlator, which is not
physical since it should imply---in unitary theories---that $\mathcal O$ is the trivial operator as a consequence of the
Reeh--Schlider theorem \cite{Haag:1992hx, Strocchi:2013awa}. The only
exception is $\Delta = \frac{1}{2}(d-2)$ which corresponds to an operator
saturating the unitarity bound; in that case we have
\begin{equation}
	\int_0^\infty\de s \, \delta(s) \frac{\widetilde C_{\frac{d-2}{2}}}{p^2+s}  = \frac{\widetilde C_{\frac{d-2}{2}}}{p^2} \ ,
\end{equation}
which is nothing but the free-theory propagator, as expected from the fact that
a scalar operator whose conformal dimension saturates the unitarity bound,
$\Delta=\frac{1}{2}(d-2)$, satisfies the equations of motion of a free scalar field. As a consequence only a single-particle state contribute to the propagator between two fundamental free scalar.

Let us now turn to case b), which is the only non-trivial possibility when
$\Delta > \frac{1}{2}(d-2)$. The integration is straightforward in position
space, and we obtain the correct expression for the two-point function of a
scalar operator of dimension $\Delta$---up to the case of the free scalar,
discussed above\footnote{The free propagator is usually normalized differently,
since we are interested in the functional dependence on the spacetime
coordinates, we omit here those factors which can be thought as
included in the definition of $\widetilde{C}_\Delta$.}
\begin{equation}
    \langle \mathcal O(x) \overline{\mathcal O}(y) \rangle = \int_0^\infty \de s \ \tilde C_\Delta s^{\Delta-\frac{d}{2}} \ G_s(x-y)  \propto \frac{\widetilde C_\Delta }{|x-y|^{2\Delta}}\,.
\end{equation}
As we recover the expression of the two-point function, this shows that the
spectral density at the CFT point---up to the free scalar field---is
the correct expression of a two-point function in CFT, confirming that the correct spectral density for a CFT is given by $\rho(s)
\propto  s^{\Delta-d/2}$.

It is instructive to study the same integral in momentum space, as it will
make the connection with conformal anomalies clear. By direct integration we
have 
\begin{equation}\label{eq:eqp}
		\langle \mathcal O(p) \overline {\mathcal O}(-p) \rangle \,=\, \int_0^\infty \de s \ s^{\Delta-\frac{d}{2}} \frac{\widetilde C_\Delta}{p^2+s} \,\propto\, \widetilde C_\Delta \,p^{2\Delta-d}\csc\left(\frac{\pi}{2}(d-2\Delta) \right)\,,
\end{equation}
Among all the factors a special role is played by the cosecant function. Indeed,
the convergence of the integral is, strictly speaking, only ensured when
$d>2\Delta$. However it is straightforward to define the analytic continuation
of the integral above so that equation \eqref{eq:eqp} holds for any $\Delta
\not \in \frac{d}{2}+\mathbb N$. In the latter cases, due to the cosecant
function, the two-point function remains divergent, and needs to be
regularized. Using dimensional regularization in $d-\epsilon$ dimensions, we
can obtain a finite result in the limit $\epsilon \to 0$ by subtracting the
divergent term. To see this, let us consider the indefinite integral
\begin{equation}
\int \de s \ \frac{s^{\Delta-\frac{d-\epsilon}{2}}}{p^2+s}  = \frac{2 s^{\Delta-\frac{d-2-\epsilon}{2}} \, {}_2F_1\left(1,\Delta-\frac{d-\epsilon-2}{2},1+\Delta-\frac{d-\epsilon-2}{2};-\frac{s}{p^2}\right)}{p^2 (2\Delta-(d-\epsilon -2))}\, .
\end{equation}
The evaluation of the above expression in the regime $s \to 0$ does not give
any divergence, however for $s \to \infty$ we have a divergent term of the form
\begin{equation}
  \left( \int_0^\infty\de s \ \frac{s^{\Delta-\frac{d-\epsilon}{2}}}{p^2+s} \right)_\text{div.} \propto  \frac{p^{2\Delta-d}}{\epsilon}\ .
\end{equation}
This divergent term needs to be subtracted with a proper counterterm: this
procedure will introduce a choice of the regularization scheme which will lead
to a constant $\tilde c$ in the two-point function and introduce a scale $\mu$.
After the subtracting the divergence, we can safely take $\epsilon \to 0$ and
the regularized two-point function reads 
\begin{equation}\label{type-B-correlator}
    \langle \mathcal O(p) \overline {\mathcal O}(-p) \rangle = p^{2\Delta-d} \left(\widetilde C_\Delta  \log \frac{p^2}{\mu^2}+ \tilde c\right) \,.
\end{equation}
This coincides with the expected result derived in references
\cite{Bzowski:2013sza, Bzowski:2015pba}.
Furthermore, in even spacetime dimensions, the operators for which the
regularization above is required have integer conformal dimension. For these,
the presence of the logarithm $\ln\mu$ is the hallmark of a type-B conformal
anomaly \cite{Deser:1993yx}, as we explain further in Section
\ref{sec:TypeBanomalies}. 

\subsection{Protected Operators Along RG Flows}

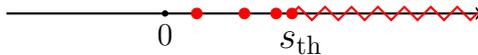
\begin{figure}[h!]
\centering
\begin{tikzpicture}
[x=0.6pt,y=0.6pt,yscale=-1,xscale=1]
\draw[->,thick] (-150,0)--(150,0);
\draw (155,-70)--(155,-55);
\draw (155,-55)--(175,-55);

\filldraw[black] (-50,0) circle (1pt) node[anchor=north]{{\color{black}{$0$}}};
\draw[thick,decorate, decoration ={zigzag},red] (30,0)--(150,0);
\draw (20,10) node [anchor=north west][inner sep=0.75pt]  [font=\large]  {$s_\text{th}$};
\filldraw[red] (30,0) circle (2pt) node[anchor=north]{};
\filldraw[red] (-30,0) circle (2pt) node[anchor=north]{};
\filldraw[red] (0,0) circle (2pt) node[anchor=north]{};
\filldraw[red] (20,0) circle (2pt) node[anchor=north]{};
\draw (160,- 70) node [anchor=north west][inner sep=0.75pt]  [font=\large]  {$s$};
\end{tikzpicture}
\caption{\emph{Schematic analytic structure of the spectral density. The red
dots indicate poles corresponding to single-particle contributions and their
location is given by the square mass of the particle exchanged. The zigzag line
denotes the branch cut corresponding to multi-particle contributions.}}
\label{PolesInSpectral}
\end{figure}

Having reviewed the spectral decomposition for CFTs, let us now apply it to the
sum rule to extract constraints on RG flows. In position space, using equations
\eqref{spectral-decomposition-Gs} and \eqref{free-propagator}, the sum rule can
be expressed as 
\begin{equation}\label{sum-rule-spec-K}
		\delta C_\Delta = 
   \frac{1}{4\Delta-2 \nu} \int \de x\  x^{2\Delta-1}\int\de s \left(x^2 s - 4\Delta(\Delta-\nu)\right) \rho(s) \left(\frac{\sqrt{s}}{x}\right)^{\nu} K_\nu(\sqrt{s}x)\,,
\end{equation}
where we used $\nu = \frac{1}{2}(d-2)$ for convenience and we used the
properties of $K_\nu(x)$ to trade the d'Alembertian for powers of $s$. It is
now crucial to know the analytic structure of the spectral density in the
complex $s$ plane. As depicted schematically in Figure \ref{PolesInSpectral},
it can be decomposed into single- and multi-particle states. The single-particle states contributions is given by
\begin{equation}
    \rho_\text{sp}(s) = \sum_i c_i \delta(s-m_i^2)   \ ,
\end{equation}
where the sum runs over all single-particle states contributing to the
two-point functions. For multi-particle states, we instead have a contribution
given by
\begin{equation}
    \rho_\text{mp}(s) = \sigma(s) \Theta(s-s_\text{th}) \ ,
\end{equation}
where $s_\text{th}$ is the threshold energy corresponding to a multiple of the
mass of the lightest exchanged particle and $\sigma(s)$ is a theory-dependent function.

Coming back to equation \eqref{sum-rule-spec-K}, we observe that
single-particle states do not contribute to the sum rule. This can be see from
the following identity:
\begin{equation}\label{single-particle-formula}
     \int \de x\, x^{2\Delta-1}\left(x^2 m_i^2 - 4\Delta(\Delta-\nu)\right)  \left(\frac{m_i}{x}\right)^{\nu} K_\nu(m_i x) = 0 \ ,
\end{equation}
for any $\Delta$ and $\nu$, and using the properties of the modified Bessel
function of the second kind $K_\nu(x)$. This is a remarkable feature of the sum
rule, and is a consequence of its specific functional form. 

Note that this assumes that the spectral decomposition can be separated in
terms of single- and multi-particle contributions, where the latter contributes
in the CFT limit at both high- and low-energy. There are indeed examples for
which this assumption is not correct, for instance in the planar-limit of
Yang--Mills theory.  This is however believed to be a large-$N$ artifact, as
multi-particle states are $1/N$-suppressed and therefore absent in the planar
limit, and not a feature of physical theories.\footnote{We thank Zohar
Komargodski for bringing this point to our attention.}

As there are no contributions from single-particle states, the sum rule can be rewritten as 
\begin{equation}\label{sum-rule-sigma}
		\delta C_\Delta = \frac{1}{4\Delta-2 \nu}\int \de  x\  x^{2\Delta-1}\int_{s_\text{th}}^\infty  ds \ \sigma(s) \left(x^2s - 4\Delta(\Delta-\nu)\right)\left(\frac{\sqrt{s}}{x}\right)^{\nu} K_\nu(\sqrt{s}x)\,,
\end{equation}
To proceed, we will use the asymptotic expansion of the spectral decomposition.
The UV divergence of the sum rule indeed fixes the leading behavior of the
spectral density to be 
\begin{equation}
    \rho(s) \sim C_\Delta^\text{UV}\frac{s^{\Delta-\frac{d}{2}}}{\Gamma(\Delta+2\nu)}  \ .
\end{equation} 
The derivation of the above asymptotic is given in Appendix \ref{eq:asymptotic
of the spectral density}. Its physical interpretation is that the spectral
density at high energy is dominated by the UV contribution of case b) in equation
\eqref{eq:CFTspectraldensity}. Assuming analyticity of the spectral density, we
find the following expansion
\begin{equation}\label{eq:rhoexpansion}
		\rho(s) =  C_\Delta^\text{UV}\,\frac{s^{\Delta-\frac{d}{2}}}{\Gamma(\Delta+2\nu)}  \left(1+\frac{a_1}{s}+\frac{a_2}{s^2}+\ldots \right)\ .
\end{equation}
This expansion is correct in free theories and can be extended to perturbation
theory, see Appendix \ref{eq:asymptotic of the spectral density} for a detailed
explanations. Since one can think of the expansion above as a (conformal) perturbation around the UV fixed point for highly non-perturbative theories (such as
RG flows starting and ending at isolated fixed points which are far away in the theory space) the expansion above can be trusted only around the UV fixed point since non-perturbative effects will arise along the RG flow.

This leads us to distinguish between two possibilities: 
\begin{itemize}
		\item[i)] The branch cut associated with multi-particle states starts
		at $s= s_\text{th}>0$. In the infrared, the protected operator
		$\mathcal{O}$ vanishes, meaning that it has a zero two-point
		function. 
		\footnote{In fact, the IR coefficient $C_\Delta^\text{IR}$ of the
		two-point function coefficient is associated with the discontinuity of
		the branch cut at low energy, i.e. $s \sim 0$.}
	\item[ii)] The branch cut instead starts from $s = s_\text{th} = 0$. This
	is the case for which the protected operator $\mathcal O$ is non-trivial in the
	infrared.
\end{itemize}
As these two cases lead to different physical behaviors, we will discuss them
separately.

\paragraph{Case i)} The branch cut starts away from the origin,
$s_\text{th}>0$. The expansion in equation \eqref{eq:rhoexpansion} does not account for non-perturbative effects, however, since single-particle states do not contributes we can use it to approximate the multi-particle contribution around the uv fixed point. The sum rule in the form given in equation
\eqref{sum-rule-sigma} is therefore 
\begin{equation}
    \delta C_\Delta = \frac{1}{4\Delta-2 \nu}\int_0^\infty dx\  x^{2\Delta-1}\int_{s_\text{th}}^\infty  ds \ \sigma(s)\ \left(x^2s  - 4\Delta(\Delta-\nu) \right) \left(\frac{\sqrt s}{x}\right)^{\nu} \mathrm K_\nu (\sqrt s x)\ .
\end{equation}
It is then easy to show by direct computation that the leading term
 in the expansion of equation \eqref{eq:rhoexpansion} is given by
\begin{equation}
    \delta C_\Delta = C_\Delta^\text{UV}\ .
\end{equation}
Moreover, it is possible to check that any other contribution will vanish. This
can be done in full generality in $d=3$, as the Bessel function simplifies to
an exponential, but is more arduous to show in higher dimensions. We have
checked this is indeed the case in various dimensions for the first few
corrections. 

Physically, our result is clear. The IR regime is encoded in the
behavior of the spectral decomposition around $s=0$. By assumption $s_\text{th}>0$ and single-particle state do not contribute, and the operator therefore vanishes in the infrared so that
$C_\Delta^\text{IR}=0$, and we obtain $\delta C_\Delta= C_\Delta^\text{UV}$.

\paragraph{Case ii)} 

The branch cut now starts at $s_\text{th}=0$, and the sum rule is given by 
\begin{equation}
	\delta C_\Delta = \frac{1}{4\Delta -2 \nu }\int_{0}^\infty  dx\  x^{2\Delta-1}\int_{0}^\infty  ds \ \sigma(s)\ \left(x^2s  - 4\Delta(\Delta-\nu) \right) \left(\frac{\sqrt s}{x}\right)^{\nu} \mathrm K_\nu (\sqrt s x)\,.
\end{equation}
The first order in the expansion given in equation \eqref{eq:rhoexpansion} can
be found using the following identity
\begin{equation}
		\int_0^\infty ds \ s^{\frac{\alpha}{2}-1}\, \mathrm K_\nu(\sqrt{s} x) = 2^{\alpha-1}x^{-\alpha} \,\Gamma\left(\frac{\alpha-\nu}{2}\right)\,\Gamma\left(\frac{\alpha+\nu}{2}\right)\,,
\end{equation}
with $\alpha>\nu$, and we obtain that at leading order the sum rule vanishes,
that is, $C_\Delta^\text{UV} \sim C_\Delta^\text{IR} $. However,
one can check that the higher-order terms diverge, although physically we
expect the difference $\delta C_\Delta$ to be finite. This means that
non-perturbative must be taken into account in order to resum the
expansion given in equation \eqref{eq:rhoexpansion}, and that $\delta C_\Delta$ is
dominated by non-perturbative effects.

Let us now summarise the two different behaviors and their physical
interpretation. At a scale $\Lambda$ along the RG flow, high-energy
contributions in the sum rule are dominated by those for which $s \gg
\Lambda$. Conversely, low-energy contributions are dominated by $s \ll
\Lambda$. In case i) the low-energy contributions are given by single-particle
states, as we except $s_\text{th}$ to be at least of the same of
magnitude of $\Lambda$. However we have found around equation
\eqref{single-particle-formula} that such single-particle states do not
contribute to the sum rule and we conclude that $C_\Delta^\text{IR}=0$ and
$\delta C_\Delta = C_\Delta^\text{UV}$.

On the other hand for case ii), massless states participate to the sum rule as
the branch cut starts at $s_\text{th}=0$, and IR contributions are non-trivial.
We have found that in that case, at leading order $C_\Delta^\text{IR}=
C_\Delta^\text{UV}$. However non-perturbative effect must be taken into
account, and we cannot conclude $\delta C_\Delta\geq0$. Despite this, when the
protected operators are associated with type-B anomalies, there is strong
evidence that this is indeed the case, as we will argue in the next paragraph.

\subsection{Type-B Conformal Anomalies}\label{sec:TypeBanomalies}

The constraints above have a natural interpretation in terms of anomaly
coefficients in even spacetime dimensions.  We will now review the connection
between conformal anomalies and the two-point function discussed above
explicitly. The standard procedure is to consider a CFT coupled to a curved
spacetime background with metric $\gamma_{\mu\nu}(x)$, and spacetime-dependent
sources $J^I(x)$ for a collection of scalar operators $\mathcal{O}_I$. The resulting
quantum effective action $W[\gamma_{\mu\nu}, J^I] = \log Z[\gamma_{\mu\nu},
J^I]$ then acts as a generating functional for connected correlator involving
the stress tensor $T_{\mu\nu}$ and the operators $\mathcal{O}_I$. We will take
the scalar operators to have dimensions $\Delta-d/2\in\mathbb{N}$. It is well
known that conformal anomalies are neatly encoded into a local anomaly obtained
from a variation of the quantum effective action $W$ under generalised Weyl
transformations \cite{Osborn:1991gm, Osborn:1991mk}
\begin{equation}
	\begin{gathered}
			\delta_\sigma W[\gamma_{\mu\nu}, J^I] = \int \de^dx \sqrt{\gamma}\,\sigma(x)\,\mathcal{A}(\gamma_{\mu\nu}, J^I)\,,\\
		\delta_\sigma \gamma_{\mu\nu} = 2\sigma(x) \gamma_{\mu\nu}\,,\qquad
		\delta_\sigma J^I(x) = -\Delta\,\sigma(x) J^I(x)\,.\qquad
	\end{gathered}
\end{equation}
An important point is that the anomaly $\mathcal{A}$ must be a \emph{local}
function of the sources and their derivatives. Up to scheme-dependent local
counterterms, one then distinguishes between two types of anomalies
\cite{Deser:1993yx}: those that vanish when integrated over spacetime at
constant $\sigma$, called type A, and those that do not, called type B. An
example of type-A anomalies is a terms of the form $\sigma a\,\sqrt{\gamma}E_d$
where $E_d$ is the Euler density, which can be written as a total derivative,
and its coefficient, $a$, is then the quantity that is relevant for the
$a$-theorem \cite{Komargodski:2011vj}. On the other hand, type-B anomalies can
also be shown to equivalently arise through an explicit $\log \mu$ dependence
in the effective action, and by extension in the associated
correlators.\footnote{Strictly speaking, type-A anomalies can also give rise to
a $\log \mu$ dependence, but only when the partition function is computed on a
space with a non-trivial topology such a the $d$-sphere, see e.g.
\cite{Gerchkovitz:2014gta, Gomis:2015yaa} and references therein.}

In the case of protected operators with integer conformal dimension $\Delta = n
- d/2$, the anomaly contains a term 
\begin{equation}
	\delta_\sigma W \supset \int \de^{d}x \, \sqrt{\gamma} \sigma (x) C_{IJ} J^I \Delta_c \overline J^J \ , \qquad \Delta_c = \Box^n + \text{curvature terms}\,,
\end{equation}
which does not integrate to zero for constant $\sigma$ and is therefore
associated with a type-B anomaly. One then finds that the anomaly coefficient is
the numerator of the two-point function by functional derivation with respect to
the sources:
\begin{equation}
		C_{IJ} = \langle \mathcal O_I(1) \overline{\mathcal O}_J(0)\rangle\propto \left.\frac{\delta^2 W}{\delta J^I(x)\delta \overline{J}^J(y)}\right|_{\overset{ x=1,y=0}{J=\text{const}}}\,.
\end{equation}
As we have seen in the beginning of this section, for a single operator, in
momentum space this correlator involves a term proportional to $\log \mu$, see
equation \eqref{type-B-correlator}, confirming the presence of a type-B
conformal anomaly.

\paragraph{Type-B anomalies along the conformal manifold}

Among Type-B anomaly coefficients, a special role is played by those related to
marginal operators, i.e. operators for which $\Delta = d$. The corresponding
Type-B anomaly coefficient
$\chi_{ij}=\left<\mathcal{O}_i(1)\overline{\mathcal{O}}_j(0)\right>$ define the
Zamolodchikov metric of the conformal manifold. The Zamolodchikov metric is
related to other type-B anomaly coefficients through an intricate web of Weyl
consistency conditions \cite{Osborn:1991gm, Osborn:1991mk, Baume:2014rla,
Niarchos:2019onf, Schwimmer:2019efk, Andriolo:2022lcb}. In particular,
it can be shown that type-B anomalies are covariantly constant with
respect to the connection $\nabla_\chi$ defined by the Zamolodchikov
metric \cite{Niarchos:2019onf,Andriolo:2022lcb}:
\begin{equation}\label{eq:WICM}
    \nabla_\chi C_\Delta = 0 \ .
\end{equation}
As a corollary, it is means that computing a coefficient at a single point of
the conformal manifold, one can in principle then find its value at any point
using equation \eqref{eq:WICM} \cite{Niarchos:2019onf,Niarchos:2020nxk}. In
particular, if it admits a free point\footnote{Depending of the particular RG flow it can be enough to assume the presence of a point of the conformal manifold in which only some of the coupling constants are small.}, computations can be made
significantly easier. Note however that while these points have a special role
in the structure of conformal manifolds \cite{Baume:2020dqd,
Perlmutter:2020buo}, they can only occur if it is non-compact
\cite{Baume:2023msm}, and their presence is therefore not guaranteed. This is
trivially the case for instance with isolated CFTs, since they do not have any
marginal deformations.

When such points exist, this is very useful to study the sum rule. Indeed,
starting from a generic point of the conformal where the spectral decomposition
on the associated RG flow is difficult to compute, we can then use
equation \eqref{eq:WICM} to go close to a free point. There the expansion given
in equation \eqref{eq:rhoexpansion} is now a good approximations, since
non-perturbative effects are expected to be suppressed. Through the arguments
above, we therefore expect that for protected operators associated with type-B
anomalies,
\begin{equation}
		\delta C_\Delta = C_\Delta^\text{UV} - C_\Delta^\text{IR} \ge 0\,.
\end{equation}
Using the covariance of $C_\Delta$ over the conformal manifold, see equation
\eqref{eq:WICM}, we can in principle use this result to compute the difference
between RG flows starting and ending at any point of the conformal manifold.
This provides strong hints that protected operators associated with type-B
anomalies indeed satisfy $\delta C_\Delta \geq0$.

Note that generically in the free-field limit we can construct protected
operators as combinations of fields which are either massless or massive, and the corresponding 
particle states play a role in the sum rule. A simple example is the case of
two scalar fields $\phi_1\,, \phi_2$ in four dimensions, where the first is
massive while the other is massless. It is then straightforward to see that the
combination $\phi_1^2+\phi_2^2$ then has $\delta C_{\Delta=2} =
\frac{1}{2}C_{\Delta=2}^\text{UV}$, since the massive scalar will be integrated out,
while the massless field will survive in the IR.  This is consistent with our
result via a direct evaluation of the sum rule, as we will see in Section
\ref{sec:examples}.  We will see explicit examples of such weak-coupling regimes
in the following section.

\paragraph{Connection with ANEC?}

The generalization of the $c$-theorem in four dimensions, the $a$-theorem,
cannot be proved by considering only two-point functions of stress energy
tensor. This is due to the fact that the $a$-coefficient itself does not appear
in the two-point function, contrary to its two-dimensional cousin. Recently a
sum rule was found for the difference $\delta a = a^\text{UV} - a^\text{IR}$
involving the three-point function of the stress tensor. Positivity of $\delta
a$ was then shown to be a consequence of that of the
expectation value of the average null energy (ANE) operator in any state
\cite{Hartman:2023qdn}. The same approach also provided an alternative proof of
the two-dimensional $c$-theorem \cite{Hartman:2023ccw}, unifying conceptually
these two theorems into a similar framework. It is then natural to ask whether the
same can be also done in the case of Type-B anomaly coefficients. 

To answer this question, let us consider the three-point function between
protected operators and the stress tensor in four spacetime dimensions \cite{Niarchos:2019onf,
Niarchos:2020nxk, Andriolo:2022lcb}:
\begin{equation}\label{eq:3ptfunction}
		\langle T^{\mu \nu}(p_1) \mathcal O(p_2) \overline{\mathcal O}(p_3)\rangle \sim\widetilde  C_\Delta \,p_2^{\Delta-2} p_3^{\Delta-2} \left(\delta^{\mu \nu}- \frac{p_1^\mu p_1^\nu}{p_1^2}\right)+\ldots  \,.
\end{equation} 
The two scalar operators can be used define the state\footnote{Another smearing factor may be necessary to to properly normalize the sum rule \cite{Hartman:2023ccw, Hartman:2023qdn}. However as it is not crucial for our discussion, we will omit this issue.} 
\begin{equation}
		\ket{\psi(p)} \simeq \int d^{4}x \ e^{i p \cdot x} \,\mathcal O(x) \ket{0}
\end{equation}  
and the stress tensor can be used to constructed the ANE operator contracting with a null vector $u$ (i.e. such that $u \cdot u = 0$) and integrating as 
\begin{equation}
    \mathcal E(v = 0,\vec x = 0) = \int_{-\infty}^\infty \de u \ T_{uu}(u,v = 0,\vec x =0) =  \int \frac{\de p_\nu \de^2 \vec p}{\pi (2\pi)^2} T_{uu}(p_u = 0, p_{v},\vec p) \ .
\end{equation}
Here we use the convention $u = -(y+i \tau)$ and $v = y-i \tau$, where $(\tau,
y, \vec x)$ are the coordinates in Euclidean signature. From the equation above
it is clear that the Type-B anomaly coefficient is proportional to a vanishing
term when we interpret the three-point function in equation
\eqref{eq:3ptfunction} as a vacuum expectation value of the ANE operator: the
tensor structure involving the momentum associated with the stress tensor $p_1$
contains either a term proportional to $\delta_{uu}$, which vanishes by
definition of a null vector, or contains $p_{1u}=0$.

\begin{figure}[h!]
    \centering
\begin{tabular}{|c|c|c|c|c| }\hline 
		dim & anomaly & Type &ANE & sum rule via 2-pt func.\\ \hline 
2 & $c$ & A & \cite{Hartman:2023ccw} & \cite{Zamolodchikov:1986gt,Cardy:1988tj} \\  \hline 
4 & $a$ & A &\cite{Hartman:2023qdn} & --- \\\hline 
4& $c$ & B & \cite{Hartman:2024xkw} & This work, \cite{Karateev:2023mrb}  \\\hline 
4&  $C_\Delta$ & B & --- &This work \\\hline 
\end{tabular}
\caption{Summary of the different anomaly coefficients in $d=2,4$, their type
		of conformal anomaly, and whether it appears in the two-point function.
		We further indicate whether a sum rule using the ANE
operator is known.}
    \label{fig:AnomaliesRecap}
\end{figure}

Note that the average null energy condition (ANEC) is a suitable tool to
describe anomaly coefficients of type A. Type-B anomaly coefficients as the
$c$-function in four dimensions and $C_\Delta$ discussed in this work are
instead more naturally defined by two-point functions at the fixed points. A sum rule for
the $c$-function in four dimensions---as opposed to the $a$-function discussed
above---was derived recently using the ANE operator in \cite{Hartman:2024xkw}.
However, contrary to the cases discussed above, the ANE operator is considered
between two different states and therefore positivity is not guaranteed. We
summarize those observations in Table \ref{fig:AnomaliesRecap}. It
would be interesting to also explore the connection between ANEC and the
anomaly coefficient $C_\Delta$ to the uniformize the results of this work with
the sum rules derived in references
\cite{Hartman:2023ccw,Hartman:2023qdn,Hartman:2024xkw}.

\section{Examples}\label{sec:examples}

In this section, we apply the sum rule given in equation
\eqref{sum-rule-position-space} to various examples, such has free fields and
supersymmetric field theories, where the constraints on type-B conformal
anomalies we have obtained in the previous section can be directly checked.

\subsection{Free Scalar Theory}\label{sec:scalar4d}

The simplest example we can consider is the free real scalar theory in $d>2$ of mass
$m$. Its action is given by
\begin{equation}
    \mathcal A = \int \de^{d}x \ \left[\frac{1}{2}(\partial \varphi)^2+\frac{m^2}{2}\varphi^2\right]\,,
\end{equation}
and the two-point between two scalars takes the simple form:
\begin{equation}\label{free-field-2pt}
    \langle \varphi(x) \varphi(0) \rangle = \mathcal N_d\left(\frac{m}{|x|}\right)^\nu \mathrm K_{\nu}(m |x|) \ ,
\end{equation}
where $\nu = \frac{1}{2}(d-2)$ and $\mathcal N_d$ is a normalization constant
we choose so that in the massless limit $m \to 0$, $\langle \varphi(x)
\varphi(0)\rangle x^{d-2} \to 1$ and such that in the ultraviolet, the
two-point function is normalized to one. One can easily check that e.g. $\mathcal N_3 = \mathcal N_5 = \sqrt{\frac{2}{\pi}}
$, $\mathcal N_4 = 1$, $\mathcal{N}_6=\frac{1}{2}$, \ldots 

The operators we are interested in are powers of the free field $\varphi$. They
satisfy our definition of protected operators as we are in a free-field theory. As discussed above, single-particle states do not contribute to the sum rule. As a consequence we expect no contribution from the sum rule when we substitute the free massive propagator in equation \eqref{free-field-2pt} and $\Delta = 2 \nu$, since only a single-particle state contributes to the spectral density. In fact one can check that this is the case by direct integration.
Let us check the sum rule for the cases $\varphi^n$ , with $n=2, 3, 4$.
The correlators are then easily computed using Wick's theorem to reduce any such
correlator in terms of the two-point function given in equation
\eqref{free-field-2pt}:
\begin{equation}
    \langle \varphi^n(x) \varphi^n(0)\rangle = w_n \langle \varphi(x) \varphi(0)\rangle^n \ ,
\end{equation}
where $w_n$ is the combinatorial factor given by Wick contractions. For the
cases we are interested we have $w_2 = 2$, $w_3 = 18$ and $w_4 = 72$. The sum
rule given in equation \eqref{sum-rule-position-space} can be now explicitly
used to compute the difference $\delta C_{n\nu}$
\begin{equation}
	\delta C_{n\nu} =\frac{2w_n \mathcal N_d^n}{\nu(2n-1)}
	\int \de |x|^2 \ |x|^{2\Delta-d} \left(|x|^2 \Box - 4\nu^2 n (n-1)\right) \left[\left(\frac{m}{|x|}\right)^\nu \mathrm K_{\nu}(m |x|)\right]^n\,.
\end{equation}

Since the infrared is a trivial theory, we expect that $\delta
C_{n\nu}=C_{n\nu}^\text{UV} =  w_d$. For $n=2,3,4$, this is checked in a
straightforward fashion by direct computation of the integral above. 

\subsubsection{The Spectral Decomposition}

In free theory, the spectral decomposition can be computed exactly since only a
certain number of multi-particle states will contribute. For the operator
$\varphi^2$, which has $\Delta = 2 \nu$, the spectral density is given by
\cite{Karateev:2019ymz}:
\begin{equation}\label{rho-free-field}
	\rho(s) = \frac{\Theta(s-4 m^2)}{N_d}
	\left |\langle 0|\varphi^2|n = 2\rangle\right|^2 \ , N_d = (2\pi)^{d-4}2^{d-1} \sqrt{s} (s-4 m^2)^{\frac{3-d}{2}} \ .
\end{equation}
The form factor can be computed straightforwardly using an oscillator
representation, and one finds
\begin{equation}
        \langle 0|\varphi^2|n = 2\rangle = \int \frac{\de^{d-1} k_1}{(2\pi)^{d-1}}\frac{1}{\omega_{\vec k_1}}\int \frac{\de^{d-1} k_2}{(2\pi)^{d-1}}\frac{1}{\omega_{\vec k_2}} \langle 0 | \op a_{p_1}\op a_{p_2}\op a_{k_1}^\dagger \op a_{k_2}^\dagger|0\rangle = 2 \ .
\end{equation}

Plugging back the form factor into equation \eqref{rho-free-field}, we conclude
that the spectral decomposition is given by
\begin{equation}
		\begin{aligned}
			\rho(s) &= \frac{4\Theta(s-4 m^2)}{(2 \pi)^{d-4}2^{d-1} \sqrt{s} (s-4 m^2)^{\frac{3-d}{2}}} \\
			& = \Theta(s-4 m^2) 2^{7-2 d} \pi ^{4-d} \left(\frac{1}{s}\right)^{3/2}\left(1+2 \frac{m^2}{s}+\mathrm O\left(\frac{m^4}{s^2}\right)\right) \ .
		\end{aligned}
\end{equation}
This expression can be used to recover the well-known expression of the two-point function of the protected operator $\varphi^2$ at any point of the RG flow. For instance, as a crosscheck in four dimension, using equation \eqref{spectral-decomposition-Gs}, evaluating the integral we have
\begin{equation}
    \langle \varphi^2(x) \varphi^2(0)\rangle = \int_0^\infty  \de s \ \rho(s) \frac{\sqrt{s}}{x} \mathrm K_1(\sqrt{s}x) = 2 \left(\frac{m}{x}\mathrm K_1(m x)\right)^2 \,,
\end{equation}
which is the correct expression in a free-field theory.

\subsubsection{Evolution Along the RG Flow}

The existence of a simple sum rule for $\delta C_\Delta$ in free-field theory enables us to understand
which energy scales contribute most by inserting an IR cutoff $\epsilon$.  For
ease of exposition, we will consider the free scalar field in $d=3$ and the
operator $: \varphi^2:$ for which $\Delta=1$ where the sum rule takes a very
simple form. We can then define the progressive contribution to the sum rule as
a function of the cutoff scale
$\epsilon$:
\begin{equation}
    \delta C_1(\epsilon) = \frac{2}{3}\int_{1/\epsilon}^\infty \de x \ x  (x^2 \Box-2) \frac{e^{-2 m x}}{x^2} = \frac{e^{-\frac{2 m}{\epsilon }} (4 m+6 \epsilon )}{3 \epsilon }\ .
\end{equation}

\begin{figure}
    \centering
    \begin{tikzpicture}
    \begin{axis}[
        xlabel={$m / \epsilon $},
        ylabel={$\delta C_1(\epsilon)$},
        domain=0:20,
        xmin=0, xmax=20,
        ymin=0, ymax=2.2,
        samples=100,
    ]
    \addplot[
        thick,
        black
    ]
     {(exp(-2/x) * (4 + 6*x)) / (3*x)};
    \addplot[
        thick,
        domain=0:20,
        red,
        dashed
    ]
    {2};
    \end{axis}
\end{tikzpicture}
\caption{Evolution of $\delta C_1(\epsilon)$ along the RG flow for he case of
	$3d$ free scalar massive theory. The quantity $\delta C_1(\epsilon)$ represents
	the progressive contribution to the difference $\delta C_1$ as a function of
	the energy scale $\Lambda=\frac{m}{\epsilon}$. 
}
    \label{fig:anomalyevolution}
\end{figure}
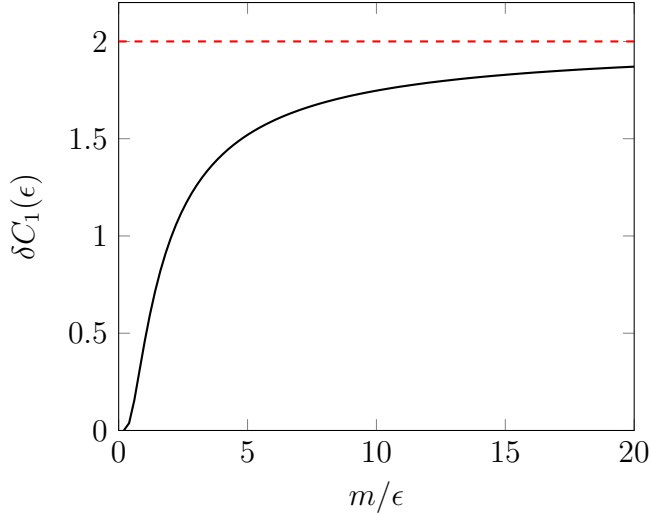

The limit $\epsilon \to \infty$ corresponds to the full sum rule, i.e. the case
where the entire RG flow---from the UV to the deep IR---is considered. The
function above is depicted in Figure \ref{fig:anomalyevolution}. One can see
that most of the information about $\delta C_\Delta$ is not encoded close to
the endpoint, but rather during the bulk of the RG flow. For generic theories,
we therefore also expect this data to be contained in the
non-perturbative regime.

In this case, the most important contributions occur as we approach the IR
fixed point ($m/\epsilon\gg 1$).
However, an expansion around
this point is not enough to have a good estimate of $\delta C_\Delta$, and
demonstrates the deeply non-perturbative nature of the sum rule. This is not
unexpected, as there is an analogous behavior in the two-dimensional
$c$-theorem \cite{Cardy:1988tj}. 

\subsection{Free Majorana Fermion in Two Dimensions}\label{sec:fermions2d}

Another natural check to perform is in the case of a two-dimensional Majorana
fermion which deformed through a mass term:
\begin{equation}
		\mathcal A = \int \de z \de \overline{z} \ \left(\psi \overline \partial \psi+\overline \psi \partial \overline \psi + i m \overline \psi \psi \right) \ ,
\end{equation}
where $\partial= \partial_z$ and $\overline \partial = \partial_{\overline z}$. 
The two-point functions are  
\begin{equation}
		\begin{gathered}
				\langle \psi(z,\overline z)\psi(0,0)\rangle = -\frac{m}{2 \pi}\left(\frac{\overline z}{z}\right)^{\frac{1}{2}}\mathrm K_1(m |z|)  \,, \qquad   
				\langle \overline \psi(z,\overline z) \psi(0,0)\rangle = i\frac{m}{2 \pi}\mathrm K_0(m |z|)  \,,\\
				\langle \overline \psi(z,\overline z)\overline \psi(0,0)\rangle = -\frac{m}{2 \pi}\left(\frac{z}{\overline z}\right)^{\frac{1}{2}}\mathrm K_1(m |z|)  \,.
		\end{gathered}
\end{equation}
We could in principle consider the operator $\overline{\psi}\psi$. However this
operator is not related to a type-B conformal anomaly, but rather one of type A. It can in fact be used to compute the central charge of the
theory, which was first performed by Cardy \cite{Cardy:1988tj} and revisited
recently in reference \cite{Hartman:2023ccw}. Instead, we will consider the
operator $(\overline \psi \psi)^2$ which is indeed related to a type-B
conformal anomaly and whose two-point correlator is given by
\begin{equation}
    \langle (\overline \psi \psi)^2(z,\overline z)(\overline \psi \psi)^2(0,0)\rangle = 2 m^4 \left(\mathrm K_0(mr)^2-\mathrm K_1(mr)^2\right)^2 \ .
\end{equation}
This operator has a conformal dimension $\Delta=2$. It is straightforward to
compute the sum rule given in equation \eqref{sum-rule-position-space} by
applying the differential operator, for which we find
\begin{equation}
    \delta C_2 = \frac{1}{8 \pi}\int \de^2 x \ x^2  \left (x^2 \Box-16 \right)  \langle (\overline \psi \psi)^2(z,\overline z)(\overline \psi \psi)^2(0,0)\rangle = 2 \,.
\end{equation}
This is the expected result, as using Wick's theory to compute the anomaly
coefficients in the UV, we find $C_2^\text{UV} = 2$; in the IR, the operator is
integrated out, so that $C_2^\text{IR} = 0$. Similar checks can be also
performed for higher powers of $\overline \psi \psi$.

\subsection{$\mathcal N = 2$ SQCD and its Higgs Phase}\label{sec:SQCD}
 
\begin{figure}[h]
    \begin{tikzpicture}
        \tikzset{square/.style={regular polygon,regular polygon sides=4, inner sep = 0}}
        \draw (0,0);
        \draw[thick] (8.3,0) node[circle,inner sep=2pt,draw] {$N$};

        \draw[thick] (8.6,-0.1) to (9.5,-0.1);
        \draw[thick,<-] (9.48,-0.1) to (10.38,-0.1);

        \draw[thick] (10.38,0.1) to (9.48,0.1);
        \draw[thick,<-] (9.48,0.1) to (8.6,0.1);

        \draw[thick] (10.8,0) node[square,inner sep=4pt,draw] {$N$};

        \draw[thick] (10.8-5,0) node[square,inner sep=4pt,draw] {$N$};

        \draw[thick,->] (8.6-2.4,-0.1) to (9.5-2.4,-0.1);
        \draw[thick] (9.48-2.4,-0.1) to (10.38-2.4,-0.1);

        \draw[thick,->] (10.38-2.4,0.1) to (9.48-2.4,0.1);
        \draw[thick] (9.48-2.4,0.1) to (8.6-2.4,0.1);

    \end{tikzpicture}
	\caption{Quiver diagram of $\mathcal N=2$ SQCD.  The circular node denotes an adjoint $\mathcal{N}=2$ vector
multiplet.
	}
    \label{fig.SCQCD}
\end{figure}
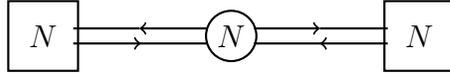

The $\mathcal N = 2$ superconformal quantum chromodynamics (SQCD) is defined by
the quiver diagram in Figure \ref{fig.SCQCD}. The gauge group $SU(N)$ of the
$\mathcal N = 2$ super-Yang--Mills theory is denoted by the central circular
node. The $U(N) \times U(N)$ subgroup of the full $U(2N)$ flavor group is
represented by square nodes. Each arrow is associated with an $\mathcal{N}=1$
hypermultiplets transforming in the bifundamental representation of the
adjacent nodes. The circular node denotes an adjoint $\mathcal{N}=2$ vector
multiplet with a scalar $\varphi$ and gaugino $\lambda$.

We will package the two scalars in the $\mathcal{N}=1$ hypermultiplets
associated with arrows flowing from left to right as $q_1$, while those
associated with arrows flowing in the other direction are denoted by $q_2$. We
follow the notation of reference \cite{Niarchos:2019onf}. They both transform
in the (anti)-fundamental representation of the flavor symmetry and the adjoint
of gauge node $SU(N)$. This enables us to RG flow by considering the specific
direction in which $q_1$ acquires a vacuum expectation value (vev), and $q_2$
does not. The full Lagrangian of SQCD is then given by 
\small \begin{equation}\label{eq:SCQDLag}
    \begin{split}
        \mathcal L  = - \operatorname{Tr} &\left[\frac{1}{4}F_{\mu \nu}F^{\mu \nu}+ i \overline \lambda_{I} \overline \sigma^\mu D_\mu \lambda^{I}+ D^\mu \varphi \overline{D_\mu \varphi}+ i g \sqrt 2\left(\epsilon_{IJ}\lambda^I \lambda^J \varphi- \epsilon^{IJ} \overline \lambda_I \overline \lambda_J\varphi \right) +\frac{g^2}{2}[\varphi ,\overline \varphi]^2\right]+\\ &+\left[D^\mu \overline q^I D_\mu q_I+ i \overline \psi \overline \sigma^\mu D_\mu \psi +i \widetilde \psi \overline \sigma^\mu D_\mu \overline{\widetilde \psi}+ i \sqrt 2 g \left(\epsilon^{IJ}\overline \psi \overline \lambda_I q_J-\epsilon_{IJ}\overline q^I \lambda^J \psi \right)\right . +\\ &+\left. f \widetilde \psi \lambda^I q_I - g \overline q^I \overline \lambda_I \overline{\widetilde \psi}+ g \widetilde \psi \varphi \psi - g \overline \psi \overline \varphi \overline{\widetilde \psi}+ g^2 \overline q_I \left(\overline \varphi \varphi +\varphi \overline \varphi\right)q^I +g^2 V(q) 
        \right]  \ ,
    \end{split}
\end{equation}
\normalsize 
where $q_I = (q_1, q_2)$ is a shorthand for the scalar of the hypermultiplets,
and $V(q)$ is the scalar potential for $q_I$, whose exact form will not be
relevant for our purpose. 

One can use the Lagrangian above to show that $\mathcal{N}=2$ SQCD is a
superconformal theory, which we take as the UV theory. We can then trigger a
specific RG flow by giving a vacuum expectation value (vev) to the scalar
$q_1$:
\begin{equation}
		\langle (q_1)_i^{a}\rangle = v \delta_{\, i}^a \,, \qquad  \langle (q_2)_i^{a}\rangle = 0 \,,
\end{equation}
where $i = 1,\dots, 2N$, $a = 1,\ldots N$ correspond to flavor and gauge
indices, respectively.

\paragraph{The (UV) CFT phase:} We will focus the anomaly coefficient
associated to the operator made out of traces of the scalar in the
$\mathcal{N}=2$ vector multiplet,
\begin{equation}
		\mathcal{O} = \operatorname{Tr} \varphi^2  \,,
\end{equation}
whose dimension is $\Delta = 2$. As the UV theory is conformal, the form of its
two-point function is fixed by symmetry; furthermore, it is a so-called
Coulomb-branch operator, and is protected by a BPS condition. A straightforward
computation using the Lagrangian in equation \eqref{eq:SCQDLag} above shows
that the two-point function takes the form 
\begin{equation}
    \langle \operatorname{Tr}\left[\varphi^2 \right](x)\operatorname{Tr}\left[\varphi^2 \right](0)\rangle = \frac{2(N^2-1)}{(2 \pi)^4} \frac{1}{|x|^4} \ .
\end{equation}
Its Fourier transform, after regularization of the divergence due to the fact that the conformal dimension of the operator is integer \cite{Bzowski:2013sza, Bzowski:2015pba, Niarchos:2019onf}, is given by 
\begin{equation}
    \langle \operatorname{Tr}\left[\varphi^2 \right](p)\operatorname{Tr}\left[\varphi^2 \right](-p)\rangle = -\frac{2(N^2-1)}{(4 \pi)^2} \log\left(\frac{p^2}{\mu^2}\right)+\tilde c \ ,
\end{equation} 
where $\tilde c$ depends on the choice of the regularization scheme, and the
presence of a logarithm $\log\mu$ confirms that we indeed have a type-B
anomaly.

\paragraph{The Higgs phase:} In the Higgs phase we can redefine the scalar
$q_1$ to take into account the vev of the field:
\begin{equation}\label{eq:shiftSQCD}
   (q_1)_{\, i}^{a}\to v \delta_{\, i}^a + (q_1)_{\, i}^a  \,,
\end{equation}
leaving $q_2$ unchanged. 
The vev furthermore induces a mass for the scalar field $\varphi$ in the vector
multiplet, $m^2 = 2g^2 v^2$, and new interaction terms which are studied in detail in reference \cite{Niarchos:2019onf,Niarchos:2020nxk}.
One of those terms can be either be computed
directly or by observing that the classical expression of the stress tensor
involves the following term
\begin{equation}
    - \frac{v}{3} \left(\partial_\mu \partial_\nu -\eta_{\mu \nu}\Box\right)\operatorname{Tr} \left[Q_1+\overline Q_1\right]\,,
\end{equation}
where $Q_1$ is the scalar associated with the top-left arrow in Figure \ref{fig.SCQCD}.
The cubic interaction between the dilaton, $\varphi$, and $\overline \varphi$
can also be obtained directly from the Lagrangian by observing that after
taking into account the redefinition given in equation \eqref{eq:shiftSQCD}, we
have 
\begin{equation}
    \mathcal L \supset -g^2 v^2 \operatorname{Tr}\left[(\overline \varphi \varphi+\varphi \overline \varphi)(Q_1+\overline Q_1)\right] \ ,
\end{equation}
where the mass term for the scalar field $\varphi$ is now explicit.
We now have all the tools needed to discuss the leading contribution to the Type-B anomaly coefficient in the Higgs phase. 

This is a particular example of the case of type-B anomalies discussed in
Section \ref{sec:TypeBanomalies}, where we start near the free-field point of the
conformal manifold. Having access to perturbation theory it is easy to devise
the fate of the anomaly coefficients of
$\mathcal{O}=\operatorname{Tr}\varphi^2$. For the purposes of the sum rule
derived in Section \ref{sec:sumrule}, since $\varphi$ obtains a mass in the
Higgs phase proportional to the vev $v$ of $q_1$, the protected operator
$\varphi^2$ is completely integrated out in the IR fixed point. From the
arguments presented in Section \ref{sec:sumruleandspectral} we expect 
\begin{equation}
		\delta C_2 = C_2^\text{UV} = 2 \frac{N^2-1}{(2 \pi)^4} \ge 0\,.
\end{equation}

\subsection{Necklace Quivers}

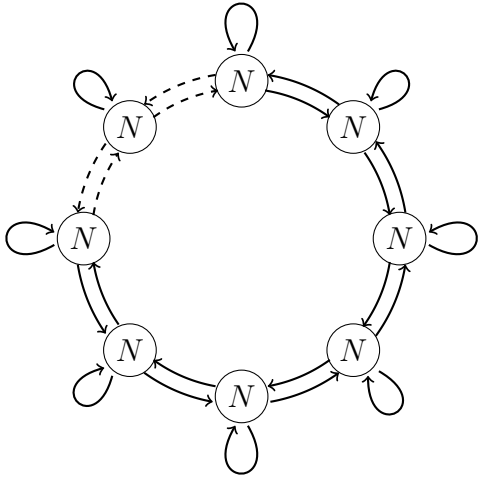
\begin{figure}
\begin{center}
\begin{tikzpicture}[scale=0.7]

\def\circledarrow#1#2#3{ 
\draw[#1,->] (#2) +(80:#3) arc(80:-260:#3);
}
 

\draw (0,3) circle [radius=0.5]  node (A) {$N$};
\draw (1.5*1.414,1.5*1.414)  circle [radius=0.5] node {$N$};
\draw (3,0)  circle [radius=0.5] node {$N$};
\draw (1.5*1.414,-1.5*1.414)  circle [radius=0.5] node {$N$};
\draw (0,-3)  circle [radius=0.5] node {$N$};
\draw (-1.5*1.414,-1.5*1.414)  circle [radius=0.5] node {$N$};
\draw (-3,0)  circle [radius=0.5] node {$N$};
\draw (-1.5*1.414,1.5*1.414)  circle [radius=0.5] node {$N$};


\draw [->,thick,domain=54:81,scale=3] plot ({1.05*cos(\x)}, {1.05*sin(\x)}) node[above right] {};
\draw [->,thick,domain=81:54,scale=3] plot ({0.95*cos(\x)}, {0.95*sin(\x)}) node[below left] {};;

\draw [->,thick,domain=9:36,scale=3] plot ({1.05*cos(\x)}, {1.05*sin(\x)});
\draw [->,thick,domain=35:9,scale=3] plot ({0.95*cos(\x)}, {0.95*sin(\x)});

\draw [<-,thick,domain=-9:-36,scale=3] plot ({1.05*cos(\x)}, {1.05*sin(\x)});
\draw [<-,thick,domain=-36:-9,scale=3] plot ({0.95*cos(\x)}, {0.95*sin(\x)});

\draw [<-,thick,domain=-54:-80,scale=3] plot ({1.05*cos(\x)}, {1.05*sin(\x)});
\draw [<-,thick,domain=-80:-54,scale=3] plot ({0.95*cos(\x)}, {0.95*sin(\x)});

\draw [->,thick,domain=54:80,scale=-3] plot ({1.05*cos(\x)}, {1.05*sin(\x)});
\draw [->,thick,domain=80:54,scale=-3] plot ({0.95*cos(\x)}, {0.95*sin(\x)});

\draw [->,thick,domain=9:35,scale=-3] plot ({1.05*cos(\x)}, {1.05*sin(\x)});
\draw [->,thick,domain=35:9,scale=-3] plot ({0.95*cos(\x)}, {0.95*sin(\x)});

\draw [->,dashed,thick,domain=-35:-9,scale=-3] plot ({1.05*cos(\x)}, {1.05*sin(\x)});
\draw [->,dashed,thick,domain=-9:-35,scale=-3] plot ({0.95*cos(\x)}, {0.95*sin(\x)});

\draw [->,dashed,thick,domain=-81:-54,scale=-3] plot ({1.05*cos(\x)}, {1.05*sin(\x)});
\draw [->,dashed,thick,domain=-54:-81,scale=-3] plot ({0.95*cos(\x)}, {0.95*sin(\x)});

\node (A1) at (0,3.35) {};
\draw[->,thick] (A1) to [out=60,in=120,looseness=15] node[above] {} (A1);

\node (A2) at (3.35,0) {};
\draw[->,thick] (A2) to [out=-30,in=30,looseness=15] node[right] {} (A2);

\node (A3) at (0,-3.35) {};
\draw[->,thick] (A3) to [out=-60,in=-120,looseness=15] node[below] {} (A3);

\node (A4) at (-3.35,0) {};
\draw[->,thick] (A4) to [out=-150,in=-210,looseness=15] node[below] {} (A4);

\node (B1) at (0.28+1.5*1.414,0.28+1.5*1.414) {};
\draw[->,thick] (B1) to [out=15,in=75,looseness=15] node[above] {} (B1);

\node (B2) at (0.28+1.5*1.414,-0.28-1.5*1.414) {};
\draw[->,thick] (B2) to [out=-30,in=-90,looseness=15] node[right] {} (B2);

\node (B3) at (-0.28-1.5*1.414,-0.28-1.5*1.414) {};
\draw[->,thick] (B3) to [out=-105,in=-165,looseness=15] node[below right] {} (B3);

\node (B4) at (-0.28-1.5*1.414,0.28+1.5*1.414) {};
\draw[->,thick] (B4) to [out=165,in=105,looseness=15] node[above] {} (B4);

\end{tikzpicture}
\end{center}

\caption{$\mathcal{N}=2$ necklace quiver. Each node corresponds to a
		$\mathcal{N}=2$ vector multiplet in the adjoint representation of
		$SU(N)$, while arrows correspond to bifundamental $\mathcal{N}=1$
		hypermultiplets.}
\label{fig:necklace}
\end{figure} 

The quiver above can be generalized straightforwardly to various types of 4d
$\mathcal{N}=2$ quiver theories. Among them, we consider the so-called necklace
quivers depicted in Figure \ref{fig:necklace}. This theory is furthermore
superconformal and we take it to be the UV CFT. There are various ways to deform this theory while preserving $\mathcal{N}=2$ supersymmetry.

In particular, we can similarly to the case of SQCD discussed above construct
Coulomb-branch operators that are protected by BPS conditions. The $i$-th node
of the quiver is associated with a $\mathcal{N}=2$ vector multiplet, which
contains a scalar $\varphi_i$. Gauge-invariant powers of this scalar are
$\mathcal{O}_i = \operatorname{Tr}\varphi^n$, as well as linear combinations
thereof are then Coulomb-branch operators forming the chiral ring of the theory.

By giving a vev to the scalar in the hypermultiplets, we can again trigger a
Higgs-branch RG flow. For cases where at least part of the chiral ring is
preserved along the flow, we can track the evolution of their two-point
function coefficients.  Each of the gauge couplings can be interpreted as a
coordinate of the conformal manifold, and as a result we can always use the
covariance of $C_\Delta$ along the conformal manifold to reach a free-field
limit of the necklace quiver and perform a computation there
\cite{Niarchos:2019onf,Niarchos:2020nxk}.  It can then be shown that only
operators of the so-called untwisted sector survive to the infrared as they remain
massless, and their coefficients is therefore the same at both endpoints:
$\delta C_{\Delta}=0$.  On the other hand, the fields of the twisted sector
become massive due to the vev of the hypermultiplets, and are therefore
integrated out in the deep IR so that $C_\Delta^\text{IR}=0$. We therefore
conclude that as expected from the discussion in Section
\ref{sec:TypeBanomalies}, in both cases $\delta C_\Delta =
C_\Delta^\text{UV}\geq0$. Similar arguments can be made for other $(\mathcal{N}=2)$-preserving deformations, such as those involving Higgs-branch operators.

\subsection{Holographic Renormalization}
\label{sec:Holo}

Even though most of the results from holography are derived in the context of
AdS/CFT, it is possible to describe RG flows by considering holographic
renormalization techniques \cite{Bianchi:2001de, Bianchi:2001kw,
DHoker:2002nbb}. Since we only focus here on two-point functions of scalar
operators let us consider the toy model defined by the  action 
\begin{equation}\label{holo-toy-action}
	\mathcal A = \frac{1}{4 \pi G}\int \de^{d+1}x\  \sqrt{g} \left[-\frac{1}{4}R+\frac{1}{2}(\partial_\mu \phi)^2+ V(\phi)\right] \,,
\end{equation}
where we demand the potential to both a maximum and a minimum. If an RG flow is
triggered on CFT living on the boundary of AdS, it corresponds in the bulk to a
flow from the maximum of the potential to its minimum. Since the two fixed
points are described by CFTs on the boundary, we need to require that at the
two stationary points the potential takes the form
\begin{equation}
      V(\phi_\text{UV/IR}) = -\frac{d(d-1)}{4 L_\text{UV/IR}^2 }\ ,
  \end{equation}
where $\phi_\text{UV/IR}$ are the value of the field $\phi$ at either or the
two stationary points, and $L_\text{UV/IR}$ is the AdS radius at these points.
Because of Poincar\'e invariance,  at the boundary the bulk metric is given by
the \textit{domain-wall Ansatz}:
\begin{equation}
	\de s^2 = e^{2A(r)} \de x_i \de x^i+ \de r^2 \ , \hspace{1 cm} \phi= \phi(r) \ .
\end{equation}
The null-energy condition and its connections with the boundary $c$-theorems is
well known \cite{Myers:2010tj}, and we now briefly review it. Using Einstein's
equation for the action defined in equation \eqref{holo-toy-action}, the warp
factor is related to the null-energy condition:
\begin{equation}
      A'' = \frac{2}{d-1} \left(T_{\, i}^{i}-T_{\, D}^{D}\right) < 0 \ ,
\end{equation}
where primed quantities are derivatives with respect to $r$. It is then
possible to define the monotonically decreasing $c$-function \cite{Myers:2010xs}
\begin{equation}
      c(r) = \frac{\pi^\frac{d}{2}}{G \ \Gamma(d/2)} \frac{1}{\left(A'\right)^{d-1}}\,,
\end{equation}
which matches the $c$-coefficients in the UV and IR CFT points since
\begin{equation}
      c_\text{UV/IR} = \frac{\pi^\frac{d}{2}}{\Gamma(d/2) G} L_\text{UV/IR}^{d-1}\ .
\end{equation}
The $c$-theorems are therefore a consequence of the fact that $L_{UV}\ge
L_{IR}$.
   
We have focused in this work on two-point functions coefficients. To compute
them in the bulk, we consider the quadratic expansion of the potential close to
the fixed points of the boundary theory, that is close to the stationary points
of the potential:
\begin{equation}
		V(\phi) \sim V(\phi_\text{UV/IR})+ \frac{1}{2}\frac{m_\text{UV/IR}^2}{L_\text{UV/IR}^2} (\phi-\phi_\text{UV/IR})^2  \ ,
\end{equation}    
where 
\begin{equation}
	m_\text{UV/IR}^2 = L_\text{UV/IR}^2 V''(\phi_\text{UV/IR})\,.
\end{equation} 
The boundary dual of $h$ is an operator whose dimension is related to the mass $m_i$ via the standard relation
\begin{equation}
       \Delta = \frac{d+\sqrt{d^2+4 m_{UV/IR}^2 L_{UV/IR}^2}}{2} \ .
\end{equation}
For non-marginal operators, the fact that the operator is protected is ensured
by the condition $m_{UV}L_{UV} = m_{IR}L_{IR} \neq 0$, since this implies that the conformal dimension in the UV is equal to the conformal dimension in the IR. Finally let us evaluate the
two-point function in the two critical points of the boundary theory. In the
AdS/CFT correspondence, since the supergravity action is proportional to $L^{d-1}$ we have
that
\begin{equation}
    \langle \mathcal O(x) \overline{\mathcal O}(0)\rangle \propto \frac{L_{UV/IR}^{d-1}}{|x|^{2\Delta}} \ ,
\end{equation}
implying, by definition of $C_\Delta$, that
\begin{equation}
       C_\Delta \propto c \propto  L^{d-1} \ .
\end{equation}
Therefore in this toy model, the positivity of $\delta C_\Delta = C_\Delta^{UV}
- C_\Delta^{IR}$ is a consequence of the classical null energy condition. The
argument above follows the procedure of \cite{Myers:2010tj} and makes use of
the classical null energy condition.

Note that is was pointed out in \cite{Nakayama:2012jv} that at the quantum
level the null energy condition can be violated, and the $c$-function must be
modified in order to be monotonically decreasing. Nonetheless here we only
require only a weaker version: $C_\Delta^{UV}> C_\Delta^{IR}$. We therefore
expect this result to hold even though suitable modifications could be required
in order to define $C_\Delta$ along the RG flow.

\section{Conclusions}\label{sec:conclusions}

In this work we have studied the evolution of two-point functions of protected
operators along RG flows. Our main results consists of a sum rule for the
two-point function coefficients $\delta C_\Delta =
C_\Delta^{UV} - C_{\Delta}^{IR}$. Even if the examples we have discussed are free
fields and supersymmetric models, the sum rule does not rely on any specific
hypothesis for the symmetries of the theory, apart from the existence of
protected operators whose conformal dimensions in the UV and IR fixed point are the same. Our derivation was also adapted to derive a sum rule for the
evolution of the central charges associated with flavor currents the stress
tensor in $d>2$. 

We then combined the sum rule with the spectral decomposition of the two-point
function, showing that single-particle states do not contribute. Furthermore
we have explicitly shown that if multi-particle states start contributing at
non-zero energies, the infrared value of the two-point function is zero, which
is expected physically. We have also explored the case in which certain
multi-particle states are massless and the associated branch cut in the
spectral decomposition starts at zero. This paper mainly use perturbative expansions around the UV fixed point. It would be interesting to
further investigate this case by including non-perturbative effects in specific
examples in the future.

We have checked our results in examples in various spacetime dimensions for
different values of the conformal dimensions of the protected operators. We have
in particular focused our attention on even spacetime dimensions and integer conformal
dimensions, as there the coefficients are related to type-B conformal
anomalies. Despite previous results for the positivity of type-A anomalies---the
$c$-theorem in two dimensions \cite{Zamolodchikov:1986gt} and the $a$-theorem
in four dimensions \cite{Komargodski:2011vj}---similar results on the positivity
of type-B anomaly coefficients have not been proved to date. Here, we have
argued that combining the sum rule we have derived, the spectral density of the
two-point functions, and Ward identities derived in references
\cite{Niarchos:2019onf, Niarchos:2020nxk, Andriolo:2022lcb}, if the UV theory
has a conformal manifold containing a free-field limit, then the quantity
$\delta C_\Delta$ is positive. We stress that in even dimensions greater than
two, although the $c$-coefficient and $C_\Delta$ are both type-B anomalies,
there are important differences. In particular, in four dimensions there are
counterexamples to a possible $c$-theorem. For instance, this is the case of
$\mathcal{N}=1$ SQCD \cite{Anselmi:1997am}. However this is not in
contradiction with our results, as in $\mathcal{N}=1$ theories the stress
tensor supermultiplet does not contain any protected scalar operators.

There are many directions that are worth exploring. As discussed in Section
\ref{sec:sumruleandspectral}, there are possible connections with average
null-energy (ANE) operators used the average null energy condition (ANEC). The
latter offering a unified proof of the $a$- and $c$-theorem,
\cite{Hartman:2023ccw, Hartman:2023qdn}. As we have shown, a direct connection
with type-B conformal anomalies seems more arduous as the analogue of the
simplest correlators used there vanish upon insertion of an ANE operator. However, it
was shown that in four dimension the $c$-coefficient can be written in terms of
a non-diagonal matrix element of the ANE operator, and it would therefore be
interesting to check whether similar arguments can be made with the coefficients
$C_\Delta$.

Another direction would also be to explore the holographic implications of the
sum rule. Recently, there has been renewed interest in defining a notion of
distance between AdS vacua, especially in the context of string
compactifications \cite{Li:2023gtt, Palti:2024voy, Mohseni:2024njl,
Debusschere:2024rmi}. As there is strong evidence that $\delta C_\Delta$ decreases
along the RG flow, it provides a coarse notion distance between two CFTs,
and it would be interesting to tackle the question from the other side of the
holographic duality through more complicated examples, particularly given that
chiral-ring operators have a natural interpretation in string theory.

Matching for type-B conformal anomalies at the endpoints of the RG flow is furthermore an open subject
of discussion in the literature \cite{Schwimmer:2010za, Niarchos:2019onf,
Niarchos:2020nxk, Karateev:2023mrb, Schwimmer:2023nzk}. It would be worth 
further studying this aspect in light of our results, in particular the
question of dilaton contributions to conformal anomalies to give an independent proof of the main results in this work. It
would also be interesting to make a connection with the tools developed in
reference \cite{Schwimmer:2023nzk}.  

The applications and examples presented in this work are the simplest and
most studied in literature; a natural direction would be to investigate and
validate our results in more involved examples. A possible direction could be
an exploration of short RG flows where conformal perturbation theory is a good
approximation \cite{Karateev:2024skm}. Furthermore our derivation is general
enough to also be used in the case of defect RG flows. When the bulk is fixed
to be a CFT, and the theory defined on the defect is free to flow, the defect
UV and IR theories are conformal but it is not along the flow.
Interesting examples of such flows have been discussed in the literature
\cite{Cuomo:2021rkm,Beccaria:2017rbe}. Finally, it would be interesting to further
study the sum rules for flavor currents and the stress tensor we have
derived.

\section*{Acknowledgements}
It is a pleasure to thank Ant\'onio Antunes, Zohar Komargodski, Lorenzo di Pietro, and Ignacio
Salazar for very useful discussions and important comments at different stages of the project.
We thank in particular Zohar Komargodski for interesting suggestions and
comments on the draft. EP is supported by ERC-2021-CoG - BrokenSymmetries
101044226. AM and EP have benefited from the German Research Foundation DFG
under Germany’s Excellence Strategy – EXC 2121 Quantum Universe – 390833306.
The work of FB is supported by the German Research Foundation through a
German-Israeli Project Cooperation (DIP) grant \emph{``Holography and the Swampland''}
and in part by the Deutsche Forschungsgemeinschaft under Germany’s Excellence
Strategy EXC 2121 Quantum Universe 390833306. FB, EP and AM thank the Collaborative Research Center SFB1624
\emph{``Higher Structures, Moduli Spaces, and Integrability''}.
AM is grateful to \textit{Spring School on Superstring Theory and Related Topics} where part of this work was presented for the first time. AM thanks the Simmons Center for Geometry and Physics and Yale University for hospitality during the final stage of this project.

\appendix

\section{Conformal perturbation theory and off-critical two-point function}\label{sec:CPToff-critical}

In this appendix, we shortly review aspects of conformal perturbation theory
that are needed to derive the sum rule in Section \ref{sec:sumrule}. These
ideas trace back to references \cite{Zamolodchikov:1987ti,
Zamolodchikov:1990bk,Mussardo:2020rxh}, and utilise that in the vicinity of a critical point of
an RG flow, correlators admit an expansion in terms of the conformal data. 

We start from the ultraviolet fixed point: close to the CFT, we can write a
formal action defining the deformation in terms of relevant operators $\Phi^I$: 
\begin{equation}
		\mathcal A = \mathcal A_\text{UV CFT} \,+\, \int \de^d x \, g_I\,\Phi^I (x) \ ,
\end{equation}
Close the to fixed point, the expansion for the two-point function then takes
the form
\begin{multline}\label{UV-expansion-app}
		\langle \mathcal O(x) \overline{\mathcal O}(0) \rangle_\text{off-crit.} \sim  \langle \mathcal O(x) \overline{\mathcal O}(0) \rangle_\text{UV}\,+\,\sum_I g_I \int \ \de^d y \ \langle \mathcal O(x) \overline {\mathcal O}(0)\Phi^I(y) \rangle_\text{UV}\,+\\+ \sum_{I,J}\frac{g_I g_J}{2} \int \ \de^d y_1\int \ \de^d y_2 \ \langle \mathcal O(x) \overline {\mathcal O}(0)\Phi^J(y_1) \Phi^I(y_2)\rangle_\text{UV}+\ldots\ .
\end{multline}
Observe that this expansion is not expected to be convergent at any value of
$g_I$, but its use is justified by the fact that here we are only interested in
the behavior of the correlation function around the critical point. We further
assumed that the operator $\mathcal O$ is protected in the sense defined in
Section \ref{sec:sumrule} and therefore by simple dimensional analysis, we have
that
\begin{equation}
    \langle \mathcal O(x) \overline{\mathcal O}(0)\rangle_\text{QFT} = \frac{C_\Delta(\chi_I)}{x^{2\Delta}} \ ,
\end{equation}
where $C_\Delta$ is a function of the dimensionless quantities
\begin{align}
    \chi_I = g_I |x|^{d-\Delta_{\Phi^I}} \ .
\end{align}
As the RG flow is triggered by the presence of the relevant field $\Phi$,
implying that $\Delta_\Phi < d$, we can combine it with the expansion in
equation \eqref{UV-expansion-app} to obtain:
\begin{equation}\label{UV-expansion-C-app}
		\begin{aligned}
			C_\Delta
			&= C_\Delta^\text{UV}+ \sum_I c_1^I g_I |x|^{d-\Delta_{\Phi_I}}+\sum_{I,J} c_2^{IJ}  g_{I} g_J |x|^{2 d-\Delta_{\Phi_I}-\Delta_{\Phi_J}}+\ldots\\
			&= C_\Delta^\text{UV}+ \sum_I c_1^I\,\chi_I+\sum_{I J} c_2^{IJ} \chi_I \chi_J +\ldots\,,
		\end{aligned}
\end{equation}
where the coefficients $c_1^I$ and $c_2^{IJ}$ can be fixed in terms of UV CFT
data from the expansion above. Note that the assumption that the operator
$\mathcal O$ is protected ensures that absence of any logarithmic terms, and
only powers of $\chi_I$ appears in the expansion

A similar expansion can be performed \emph{mutatis mutandis} around the
infrared fixed point. There, we have the formal expansion
\begin{equation}
		\mathcal A = \mathcal A_\text{IR CFT}\,+\,\int d^{d}x \,\lambda_I\, \Psi^I(x)+\ldots  \ , 
\end{equation}
where we the deformation operators are now irrelevant, as in the IR those are
the types of operators we are interested in for our purpose. The two-point
function then has a similar expansion in terms of the IR conformal data:
\begin{multline}
    \langle \mathcal O(x) \overline{\mathcal O}(0) \rangle_\text{off-crit.} \sim  \langle \mathcal O(x) \overline{\mathcal O}(0) \rangle_\text{IR}+\sum_{I} \lambda_I \int \ \de^d y \ \langle \mathcal O(x) \overline {\mathcal O}(0)\Psi^I(y) \rangle_\text{ir}+\\+ \sum_{I J}\frac{\lambda_I \lambda_J}{2} \int \ \de^d y_1\int \ \de^d y_2 \ \langle \mathcal O(x) \overline {\mathcal O}(0)\Psi^I(y_1) \Psi^J(y_2)\rangle_\text{ir}+\ldots\ .
\end{multline}
Combining this expansion with the fact that the operator is protected implies
that
\begin{equation}
		C_\Delta  = C_\Delta^\text{IR}+\sum_I\tilde{c}_1^I\, \lambda_I |x|^{d-\Delta_{\Psi^I}}+\sum_{I,J} \tilde{c}_2^{IJ} \lambda_I \lambda_J |x|^{2d-\Delta_{\Psi^I}-\Delta_{\Psi^J}}+\ldots  \ .
\end{equation}
The main difference with the expression given in equation
\eqref{UV-expansion-C-app} is that now $\Delta_{\Psi^I} >d$ as $\Psi^I$ are
irrelevant operators. Note that in general the number of irrelevant
perturbations can be infinite, however this will not be important for the
purposes of this work.

\section{The Spectral Decomposition}\label{sec:spectral-decomposition}

We review here the derivation of the K\"all\'en--Lehmann spectral decomposition
based on the one used in Appendix A of reference \cite{Cappelli:1990yc}. The
idea is based on the following steps:
\begin{itemize}
    \item[$\star$] find a basis of the Hilbert space;
    \item[$\star$] use the basis of the Hilbert space to construct a resolution of the identity;
    \item[$\star$] insert the resolution of the identity in the two-point function.
\end{itemize}

Let us consider the Fock basis of the theory spanned the single- and
multi-particle states of the theory. In an interacting theory the existence of
this basis is guaranteed at least asymptotically by the Haag--Ruelle theorem
\cite{PhysRev.112.669} assuming the existence of a mass gap. Although we only
need the existence of a basis of the Hilbert space of the theory diagonalizing
the Laplace operator---or equivalently the momentum operator $\op P_\mu$ 
---we
will make this assumption. We will denote those states by $\ket{\alpha}$.
Assuming this is a complete basis of the Hilbert space, we can write the
resolution of the identity
\begin{equation}\label{eq-app:resolution-identity}
    \op 1 = \sumint \ket{\alpha}\bra{\alpha}\  \de \alpha .
\end{equation}
We now consider a two-point function
\begin{equation}
		\langle \mathcal O(x)\overline{ \mathcal O}(0)\rangle =  \langle \mathcal O(x)\,\op 1 \,\overline{ \mathcal O}(0)\rangle  \ .
\end{equation}
Using equation \eqref{eq-app:resolution-identity}, we can rewrite the two-point
function in terms of the basis of the Hilbert space $\{\ket{\alpha}\}$:
\begin{equation}
     \langle \mathcal O(x)\overline{ \mathcal O}(0)\rangle = \sumint e^{-i p_\alpha \cdot x} \left |\langle 0|\mathcal O|\alpha\rangle\right|^2\ \de \alpha \,,
\end{equation}
where we made use of the assumption that we have chosen an eigenbasis of the
momentum operator, and used $\langle 0|\mathcal O|\alpha\rangle = \langle
\alpha |\overline{\mathcal O} |0\rangle$. The spectral density $\rho(p^2)$ is
then defined as:
\begin{equation}
    \rho(p^2) (2 \pi)^{-d} =  \sumint \delta(p-p_\alpha) \left |\langle 0|\mathcal O|\alpha\rangle\right|^2\ \de \alpha \,.
\end{equation}  
The two-point function can then be written as
\begin{equation}
    \langle \mathcal O(x)\overline{ \mathcal O}(0)\rangle= \int \frac{\de^dp}{(2 \pi)^d}\ \int_0^\infty \de s  \  \rho(s) e^{-i p\cdot x}\delta(p^2-s)  \,.
\end{equation}
Note that in the previous step, it is crucial to use the fact that $p^2 >0$.
Assuming the convergence of the two integrals, we recognize the propagator
$G_{s}(x)$ of a free scalar field of mass $\sqrt{s}$, and we conclude that
\begin{equation}
     \langle \mathcal O(x)\overline{ \mathcal O}(0)\rangle = \int_0^\infty \de s \ \rho(s) \ G_{s}(x)\,,
\end{equation}
Equivalently, in momentum space we have the decomposition
\begin{equation}
    \langle \mathcal O(p) \overline{\mathcal O}(-p) \rangle = \int_0^\infty \de s \  \frac{\rho(s)}{p^2+s} \,.
\end{equation}
The physical interpretation is quite natural: the two-point function is
expanded in terms of contribution of the element of the Fock space
$\ket{\alpha}$. Each element contributes with a free propagator weighted by a
form factor $\left |\langle 0|\mathcal O|\alpha\rangle\right|^2$. The
single-particle contributions are expected to appear in a discrete sum defining
the spectral density $\rho$ from $M^2<s<4 M^2 = s_\text{th}$, where $M$ is the mass gap of
the theory

\footnote{More precisely $M$
is not the mass gap of the theory but it is the energy of the first single
particle excitation with non-vanishing form factor.} Each of these
contributions gives a pole in the complex momentum plane. Beyond $s_\text{th}=4 M^2$ 
multi-particle states are expected to appear and form a continuum. A branch cut
is therefore expected beyond this threshold value of the momentum and we obtain
the pole structure depicted in Figure \ref{PolesInSpectral}.\footnote{In some
cases, two-particle states cannot contribute due to symmetries setting
the associated form factors to zero. The first multi-particle state is
then a three-particle state appearing at $m^2 = 9 M^2$.} 
There is also the possibility in which multi-particle states contributes from $s = 0$, for instance because of the presence of single-particle states. This is maybe the most interesting case since it is the case in which the operator $\mathcal O$ does not vanish in the infrared. In the latter case around $s = 0$ the branch-cut is dominated by the IR fixed point as the limit $s \to \infty$ is dominated by the UV fixed point.

\subsection{The Stress Tensor in CFT}\label{sec:stressTensorCFT}

We review here the application of the spectral decomposition to the stress
tensor in CFT. In general the spectral decomposition for tensors have to be
split in two components, since there are two possible Lorentz structures
corresponding to spin-$0$ and spin-$2$ states, respectively
\cite{Cappelli:1990yc}:
\begin{equation}
    \langle T_{\mu \nu}(x) T_{\rho \sigma}(0) \rangle = \sum_{J = {0,2}}\int_0^\infty\de s \  \rho^{(J)}(s) \Pi_{\mu \nu \rho \sigma}^{(J)} G_s(x) \ ,
\end{equation}
    where \begin{equation}
        \Pi_{\mu \nu \rho \sigma}^{(0)} = \frac{S_{\mu \nu}S_{\rho \sigma}}{\Gamma(d)} \ , \hspace{1 cm} \Pi_{\mu \nu \rho \sigma}^{(2)}=\frac{1}{\Gamma(d-1)}\left[\frac{d-1}{2}S_{\mu(\rho}S_{\nu \sigma)}-S_{\mu \nu}S_{\rho \sigma}\right]\ ,
    \end{equation}
$G_{m^2}(x)$ is the free massive propagator and $S_{\mu \nu} = \partial_{\mu
}\partial_\nu- \delta_{\mu \nu}\Box$. Note that the case $d= 2$ is simpler
since the term corresponding to spin-2 states is trivial and only spin-$0$
states contribute. In that case, scale invariance implies only two possibility
for the spectral function
\begin{equation}
    \text{i)} \ \rho^{(0)}(s) =\tilde c \frac{\delta(s)}{s} \ , \hspace{1 cm} \text{ii)} \ \rho^{(0)}(s) = \tilde c \frac{1}{s} \ .
\end{equation}
Observe that the first possibility is actually non-zero and non-divergent. In
fact, we can compute the correlation of the trace of the stress tensor and we
get 
\begin{equation}
    \langle T_{\, \mu}^{\mu}(x) T_{\, \rho}^{\rho}(0)\rangle \propto - \tilde c\  \Box \delta^{(2)}(x) \ .
\end{equation}
The possibility ii) gives a divergence in the correlator in position space
which is not expected and we therefore conclude that ii) is unphysical. The
conclusion is that, since the correlation function above is ultralocal, by
invoking the Reeh--Schlider theorem \cite{Strocchi:2013awa, Haag:1992hx} we
have that $T_{\, \mu}^{\mu} = 0$ and therefore scale invariance, together with
locality (existence of the stress tensor) and unitarity, implies conformal
invariance in two-dimensions. Above two dimensions we have that 
\begin{equation}
    \text{i)} \ \rho^{(J)}(s) =\tilde c\  s^{\frac{d}{2}-1} \delta(s) \ , \hspace{1 cm} \text{ii)} \ \rho^{(J)}(s) = \tilde c\  s^{\frac{d}{2}-2} \ .
\end{equation}
The case i) gives zero correlation functions for a CFT. While it is possible to
give a meaning of this type of behavior, see the discussion in Section 3 of
\cite{Cappelli:1990yc}, on the other hand the case ii) above two dimensions is
not divergent and therefore we can not exclude it. Nonetheless, the momentum
space of the two-point function of $T_{\, \mu}^{\mu}$ is naively divergent, and
after regularization a logarithmic term in momentum space appears. In even
spacetime dimensions this matches the expectation for type-B conformal
anomalies. Conversely, a type-A anomalies can appear in the two-point function
of $T_{\, \mu}^{\mu}$ only in two dimensions, since it is the only case for
which the two-point function is ultralocal. 

In higher dimensions conformal invariance is not a consequences of unitarity,
scale invariance, and locality as in two dimension---at least not the way we
have argued for the $d=2$ case.

\subsection{Asymptotic of the Spectral Density}\label{eq:asymptotic of the spectral density}

A way to recover the leading term of the spectral density is to use Tauberian
theory. In the small-$x$ limit, we recover the two-point function of the UV theory
\begin{equation}
    \int_0^\infty  \de s \ \rho(s) \left(\frac{\sqrt s}{x}\right)^{\nu}\mathrm K_\nu(\sqrt{s} x) \overset{x \to 0}{\sim} \frac{C_\text{UV}}{x^{2\Delta}} \ ,
\end{equation}
where $\nu = (d-2)/2$. Using that for large values of $x$ 
\begin{equation}
		K_\nu(\sqrt s x) \sim \sqrt{\frac{\pi}{2}} \frac{e^{-\sqrt{s} x}}{\sqrt x}\,,
\end{equation} 
one can use an inverse Laplace transform to obtain 
\begin{equation}
    \rho(s) \sim \frac{C_\text{UV}}{2 \Gamma(2 \Delta-2\nu)} s^{\Delta-\frac{d}{2}} \ ,
\end{equation}
which matches the expectation that at high energy the spectral density
reproduces the high-energy behavior of the correlator. To make the statement
mathematically precise one should invoke Tauberian theorems, see references 
\cite{tauberian} for a review and
\cite{Vladimirov:1978xx,Mukhametzhanov:2019pzy,Pal:2019zzr,Pal:2023cgk,Qiao:2017xif,Marchetto:2023xap,Ganguly:2019ksp,Pal:2020wwd}
for examples of applications in physics. The main obstacle in doing this is
the presence of the Bessel function. In three spacetime dimensions the Bessel
function reduces to a simple exponential and the Tauberain theorem follows. In
$d \neq 3$ however the proof require some modification. In $d = 2$ for instance
the kernel is simply given by $\mathrm K_0$ which corresponds to the asymptotic
of the one-dimensional conformal blocks. In the latter case a Tauberian theorem
was proved in reference \cite{Qiao:2017xif}. It would be interesting to adapt the proof
to any spacetime dimension. If we add the assumption that the spectral density
is an analytic function we can simply recover the expansion
\begin{equation}
    \rho (s) = \frac{C_{uv}}{2 \Gamma(2 \Delta-2\nu)} s^{\Delta-\frac{d}{2}} \left(1+\frac{a_1}{s}+\frac{a_2}{s^2}+\ldots \right)\ .
\end{equation}
The motivation for such an assumption in a generic, non-Lagrangian theory is unfortunately not provided. It is however possible to show that this is correct in free theories and perturbation theory. In fact one can show that the spectral decomposition is related with the imaginary part of the correlator or its discontinuity\begin{equation}
    \rho(s) \sim  \operatorname{Im} \tilde G(k)  \sim  \operatorname{disc} \tilde G(k) \ ,
\end{equation}
which satisfy the analyticity condition above in perturbative examples. Clearly non-perturative effects are not completely captured by the exapansion above.

\footnotesize

\bibliography{Draft.bib}
\bibliographystyle{utphys}

\end{document}